\documentclass[12pt]{article}
\usepackage{jheppub}
\usepackage{graphicx}
\usepackage{relsize}
\usepackage{tikz}
\usepackage{amsthm,amsmath,amssymb,bbm}
\usepackage{dsfont} 
\usepackage{enumitem} 
\usepackage{relsize}
\usepackage[utf8]{inputenc}
\usepackage{relsize}
\usepackage[T1]{fontenc} 
\usepackage{lmodern}
\usepackage{caption}
\usepackage{floatrow}
\usepackage{dcolumn}
\usepackage{bm}
\usepackage{color}
\usepackage[normalem]{ulem}
\usepackage{subfig}
\usepackage{mathtools}
\usepackage{scalerel,stackengine}
\usepackage{xcolor}

\makeatletter

\def\@fpheader{\relax}

\makeatother

\newcommand{\be}{\begin{equation}}
\newcommand{\ee}{\end{equation}}
\newcommand{\bea}{\begin{eqnarray}}
\newcommand{\eea}{\end{eqnarray}}

\newcommand{\ket}[1]{| #1 \rangle}
\newcommand{\bra}[1]{\langle #1 |}
\newcommand{\scalar}[2]{\left\langle #1 | #2 \right\rangle}

\newcommand{\mean}[1]{\left\langle #1 \right\rangle}

\DeclareMathOperator{\Tr}{Tr}

\title{Cosmological states in loop quantum gravity on homogeneous graphs}

\author[a,c]{Bekir Bayta\c{s},}
\emailAdd{bekirbyts@gmail.com}
\author[b]{Nelson Yokomizo}
\emailAdd{yokomizo@fisica.ufmg.br}

\affiliation[a]{Department of Physics, {\.{I}}zmir Institute of Technology, G{\"{u}}lbah{\c{c}}e, Urla 35430, {\.{I}}zmir, Turkey}
\affiliation[b]{Departamento de F\'isica - ICEx, Universidade Federal de Minas Gerais, CP 702, 30161-970, Belo Horizonte - MG, Brazil}
\affiliation[c]{Center for Relativity and Gravitation, Beijing Normal University, Beijing, China}

\abstract{
We introduce a class of states characterized by proposed conditions of homogeneity and isotropy in loop quantum gravity and construct concrete examples given by Bell-network states on a special class of homogeneous graphs. Such states provide new representations of cosmological spaces that can be explored for the formulation of cosmological models in the context of loop quantum gravity. We show that their local geometry is described in an automorphism-invariant manner by one-node observables analogous to the one-body observables used in many-body quantum mechanics, and compute the density matrix representing the restriction of global states to the algebra of one-node observables. The von Neumann entropy of this density matrix provides a notion of entanglement entropy of a local region which respects automorphism-invariance and can be applied to states involving superpositions of distinct graphs.
}

\begin{document}

\maketitle
\flushbottom

\newpage


\section{Introduction}

Quantum effects in gravitation are expected to play a significant role in the dynamics of the Universe at times sufficiently far in the past for the energy density to reach the Planck scale. In this regime, quantum corrections to the Friedmann equation should become relevant, leading to modifications of the dynamics of the standard cosmological model. Possible signals of the dynamics of the Planck era in cosmological observations have been investigated in models of quantum cosmology as a potential path for the observation of quantum gravity phenomena \cite{amelino-camelia,AAN,barrau-rev}. Models of quantum cosmology are also explored for the analysis of conceptual and foundational question in quantum gravity in the simpler context of highly symmetric spacetimes \cite{kiefer,bojowald-qc,AS}.

The general strategy pursued in quantum cosmology is the quantization of a classical minisuperspace \cite{kiefer,bojowald-qc}. After restricting to a special class of highly symmetric classical spacetimes, the remaining global degrees of freedom of the symmetry-reduced model are then quantized. If one restricts to homogeneous and isotropic spaces, for instance, the geometry is described by a single degree of freedom---the scale factor, evolving according to the Friedmann equation---which can be quantized for the formulation of a quantum cosmology. In the case of loop quantum cosmology (LQC), the geometry is quantized with loop inspired techniques \cite{AS,ABL}.

A robust prediction of LQC is the resolution of the Big Bang singularity, which is replaced in the model by a bounce connecting a contracting universe to our expanding universe through a Planck era where the gravitational interaction becomes effectively repulsive due to quantum gravity effects \cite{AS}. The model also leads to the prediction of quantum gravity corrections to the CMB spectra \cite{AAN}, which can in particular alleviate tensions observed between the predictions of the $\Lambda$CDM model and CMB data \cite{agullo-2021,AGJS}.

It is natural to ask whether the picture provided by LQC can be recovered as some effective description of the dynamics of symmetric spacetimes in the full theory of loop quantum gravity. This would clarify how the dynamics of the symmetry-reduced model captures the evolution of symmetric states in the full theory and under what conditions its predictions can be trusted. Alternative formulations of the dynamics of loop quantum gravity (LQG) have been explored with this purpose in view, as quantum reduced loop gravity \cite{qrlg-1,qrlg-2}, spinfoams \cite{spinfoam-cosmology-1,spinfoam-cosmology-2} and group field theory \cite{gft-cosmology-1,gft-cosmology-2, gft-cosmology-3}. The results obtained so far present indications that a bouncing cosmology as described by LQC might be recovered, but a clear correspondence has not yet been reached.

Quantum reduced loop gravity, for instance, leads to a cosmology described by a so-called emergent-bouncing Universe \cite{qrlg-1,qrlg-2}, which agrees with the post-bounce, but not with the pre-bounce evolution of LQC. In the group field theory approach, the effective dynamics can describe a bouncing cosmology, which is not identical to that described by LQC, however, and other possibilities remain viable depending on how parameters involved in the construction of the model are chosen \cite{gft-cosmology-1,gft-cosmology-2}. Spinfoam cosmological models were shown to have a correct classical limit \cite{spinfoam-cosmology-1,spinfoam-cosmology-2}, but are not sufficiently developed for the evaluation of quantum corrections to the classical dynamics required for a comparison with the predictions of LQC. The embedding of LQC into the full theory of loop quantum gravity remains thus an open problem. The analysis of this problem in the covariant approach of spinfoam dynamics is of particular interest.

In order to study the dynamics of a quantized cosmological spacetime of LQC in the framework of LQG, two main questions must be addressed. First, one needs to find an adequate representation in LQG for the semiclassical homogeneous and isotropic geometries described by the effective equations of LQC. Next, given such a representation, one must compute the evolution of the relevant geometric properties. In analogy with classical cosmology, the first question corresponds to the identification of a quantum version of FLRW spaces. As LQC describes an effective evolution of classical geometries, one must also identify the relevant semiclassical states within the family of homogeneous and isotropic quantum geometries. The second question is analogous to the derivation of the Friedmann equation from the Einstein field equations.

\begin{figure}[htbp]
\includegraphics[scale=0.5]{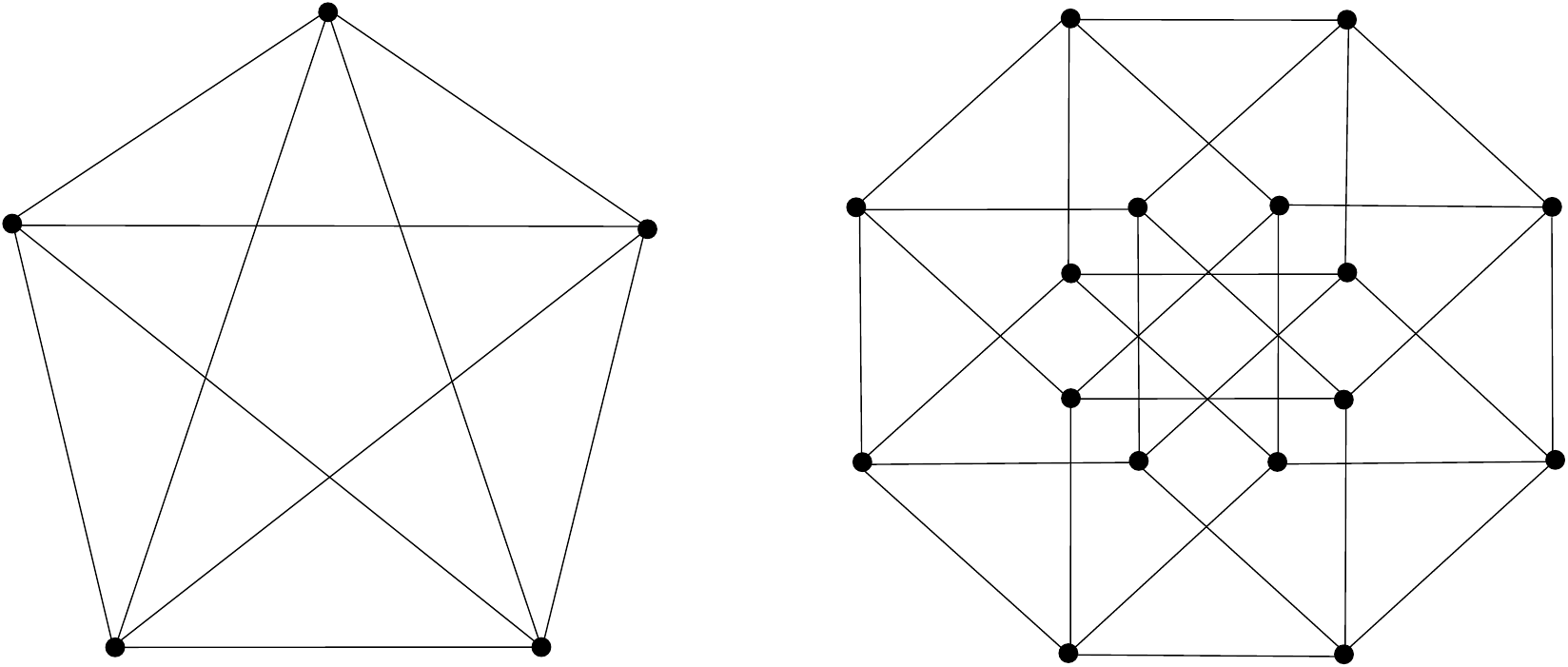}
\caption{Examples of $2$-CH graphs. \emph{Left:} The pentagram graph $K_5$. \emph{Right:} The hypercube graph $Q_4$.}
\label{fig:2-CHgraph-intro}
\end{figure}

In this work, we are concerned with the first problem regarding the construction of states representing cosmological spaces in LQG and the identification of semiclassical states among them. We first introduce an abstract class of quantum states of the geometry characterized by symmetry conditions that mirror the properties of homogeneity and isotropy of classical cosmological spacetimes, which we call cosmological states $\ket{\Psi} \in \mathcal{K}$. For such states, all nodes are equivalent, as well as all links at each node. They are defined as superpositions of automorphism-invariant states on so-called $2$-CH graphs $\Gamma_C$,
\begin{equation}
\mathcal{K} = \bigoplus_{\Gamma_C} \, \mathcal{K}_{\Gamma_C} \,,
\end{equation}
where $\mathcal{K}_{\Gamma_C}$ is the space of gauge- and autormophism-invariant states on $\Gamma_C$. In a $2$-CH graph, for any two pairs of adjacent nodes $n_1 \sim n'_1$ and $n_2 \sim n'_2$, there is an automorphism $A \in \mathrm{Aut}(\Gamma_C)$ such that
\be
A(n_1) = n_2 \quad \mathrm{and} \quad A(n'_1) = n'_2 \, .
\ee
As a result, the quantum geometry of automorphism-invariant states on $2$-CH graphs does not distinguish among the nodes or links of the graph. Examples of connected $2$-CH graphs are represented in Fig.~\ref{fig:2-CHgraph-intro}.

The cosmological states $\ket{\Psi} \in \mathcal{K}$ generalize the cases of a coherent state peaked on a cubic lattice over a cubulation, as used in quantum reduced loop gravity, or on a gluing of two regular tetrahedra, as used in spinfoam cosmology. We show that a new family of concrete examples of cosmological states are given by Bell-network states \cite{bellnetwork} on $2$-CH graphs. Bell-network states are obtained by maximizing the correlations between the fluctuations of the geometry for neighboring regions \cite{bellnetwork}. On a $2$-CH graph, this can be done in a uniform way, allowing the construction of states satisfying the required symmetry conditions for a large class of graphs. These provide a new set of candidates for the analysis of the evolution of the effective geometry of cosmological spacetimes in LQG.

In the combinatorial definition of the space of states of LQG, it is natural to require physical states of the geometry to be invariant under the graph automorphisms \cite{rovellicombi}. This condition ensures that the definition of the states depends only on the combinatorial structure of the graph, and not on the choice of a particular presentation, and can be seen as a discrete analogue of diffeomorphism invariance \cite{arrighi,colafranceschi}. More specifically for our purposes, the invariance under autormorphisms is crucial for the states to satisfy the proposed conditions of homogeneity and isotropy. Autormophism-invariance severely restricts the space of states and observables on the class of highly symmetric $2$-CH graphs explored for the definition of cosmological states.

We develop a basic set of tools for the study of homogeneous states in an automorphism-invariant way. We show that local properties of the geometry are captured by one-node observables analogous to the one-body observables used in many-body quantum mechanics \cite{martin-rothen}. We obtain such one-node observables $\mathcal{O}^1_{\mathrm{inv}}$ through the application of a group-averaging procedure to noninvariant observables $\mathcal{O}_n$ that act on a specific node:
\be
\mathcal{O}_n \,\,\,  \longrightarrow \,\,\,\, \mathcal{O}^1_{\mathrm{inv}} = \frac{1}{|\mathrm{Aut}(\Gamma)|} \,\, \sum_A \, U_A \, \mathcal{O}_n \, U_A^{-1} \,,
\ee
where $U_A$ is the representation of an automorphism $A$. The definition of the one-node observables can be extended to homogeneous states involving a superposition of states on distinct graphs. We derive an explicit formula for the density matrix $\rho_C$ representing the restriction of a cosmological state $\ket{\Psi}$ to the algebra of invariant one-node observables $\mathcal{O}_C$ on $\mathcal{K}$. The density matrix $\rho_C$ characterizes the properties of the quantum geometry as observed at a single node, whose location remains indefinite, however, as required by automorphism-invariance on homogeneous graphs.

A simple criterion for the identification of semiclassical states of the geometry in any theory of quantum gravity has been proposed in \cite{bianchi-myers}. It is argued there that, for semiclassical states of the geometry, the entanglement entropy of a bounded region of space must be proportional to the area of its boundary. In order for this proposal to be applicable to the case of cosmological spaces in LQG, a precise definition of the geometric entanglement entropy is required. This is not trivial for two reasons. First, the space of physical states does not admit a factorization into a tensor product of local Hilbert spaces associated with specific regions of space. In fact, all nodes are equivalent for homogeneous graphs. Second, it is not immediately clear how to identify a local region when the state of the geometry includes a superposition of distinct graphs.

We show that these obstructions can be overcome if a local region is specified by the observation of a boundary geometry; for instance, by the observation of some boundary spins $\{j_a\}$. Under the condition that a boundary region has been observed, measurements performed within such a region can be described by invariant observables for general states that may include a superposition of graphs. When the local region corresponds to a single node, it is characterized by the restriction $\rho_C$ of the state of the geometry to the algebra of one-node observables. In this case, we define the geometric entanglement entropy as the von Neumann entropy of such a density matrix,
\be
S_C = - \Tr ( \rho_C \log \rho_C) \,.
\ee
This provides a strategy for the calculation of the geometric entropy of cosmological states that can be applied to check whether an area law is satisfied or not. The construction can be extended for larger local regions including more than one node. In order to illustrate the application of the method, we compute the entanglement entropy of a node and of a region formed by two adjacent nodes for a cosmological state involving a superposition of Bell states on distinct graphs.

This manuscript is organized as follows. In Section \ref{sec:graphs}, we review results on symmetric graphs in order to fix the notation and introduce the connected-homogeneous graphs used later in the paper for the construction of homogeneous and isotropic states in LQG. In Section \ref{sec:homogeneity-isotropy-lqg}, we discuss automorphism-invariant states and observables on arbitrary graphs, and show that certain Bell-network states are automorphism-invariant in all graphs. Then we focus on the cases of $1$-CH and $2$-CH graphs and show that the symmetries of invariant states on such graphs can be interpreted as discrete versions of homogeneity and isotropy, respectively. A space of homogenous and isotropic states that allows for the superposition of graphs is introduced at the end of the section. In Section \ref{sec:entropy}, we define the entanglement entropy of a local region for homogeneous states and compute it for an example involving a superposition of Bell-network states on distinct graphs. In Section \ref{sec:discussion}, we summarize the results and discuss possible further developments of the techniques introduced in this work.


\section{Graph theory and notions of symmetry}
\label{sec:graphs}

A general graph $\Gamma$ is a pair $(N(\Gamma), L(\Gamma))$, consisting of a set of nodes $N(\Gamma)$ and a set of links $L(\Gamma)$, where a link $\ell \in L(\Gamma)$ is a pair of nodes. The order $N=|N(\Gamma)|$ and size $L=|L(\Gamma)|$ of $\Gamma$ are the number of nodes and links in $\Gamma$, respectively. If a link $\ell$ includes the node $n$, we say that the link is attached to this node. Ordering the links attached to a node $n$, we can denote the $a$-th link attached to it by $\ell = (n,a)$ or $\ell=na$. A directed graph is a graph equipped with an ordering of the nodes of each link. A link with ordered nodes is called an oriented link or an arc.

\begin{figure}[htbp]
 \includegraphics[scale=0.5]{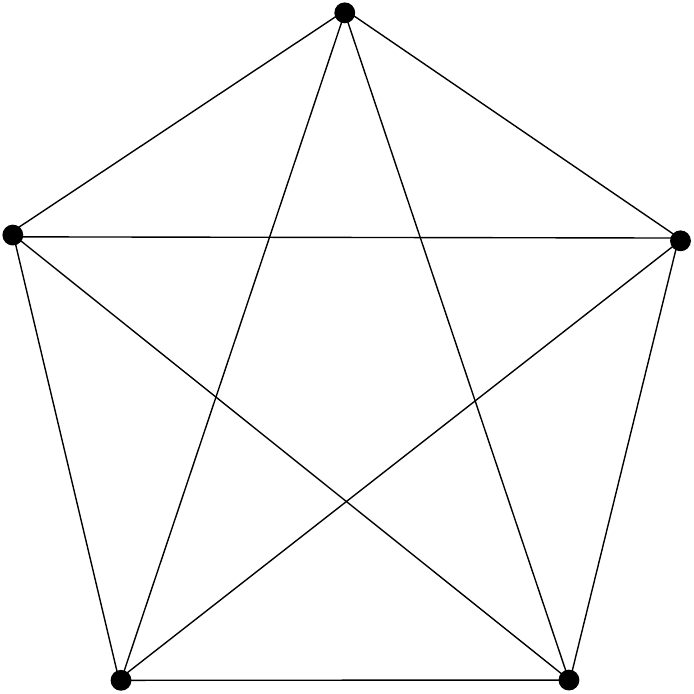}
\caption{The complete graph $K_5$ (or pentagram), with $N=5$ and $L=10$. All nodes have the same valency, $V= 4$.}
\label{fig:pentagram}
\end{figure}

A graph is simple if it has no self-loops (a link connecting a node to itself) and no multi-links (two or more links connecting the same pair of nodes); otherwise, the graph is called a multi-graph. The size of a simple graph is bounded by the order of the graph, $L \leq N (N+1)/2$. An example of a simple, regular graph is the complete graph $K_5$ depicted in Fig.~\ref{fig:pentagram}.

Let us list some basic definitions used in graph theory in order to fix our notation \cite{bondy08,harary72,biggs}. A list of graphs, including those relevant for the classification theorems that will be described in this section, is presented in the Appendix \ref{sec:graph-list}.

\begin{itemize}
\item \label{cond1}  Two nodes $n,n'$ are adjacent if they belong to a common link. The relation of adjacency is represented by $n \sim n'$. The nodes $n,n'$ of a link $\ell$ are called the endpoints of the link $\ell$.
\item \label{cond2} The adjacency matrix $\mathcal{A}$ is defined as the $N \times N$ matrix with elements $\mathcal{A}_{nn'} = 1$, if $n \sim n'$, and $\mathcal{A}_{nn'} = 0$, otherwise. It describes the connectivity of the graph.
\item \label{cond3}  The total number of links attached to the node $n$ is the valency (or degree) $V_{n}$ of the node. The sum of the valencies of all nodes in the graph is twice the size of the graph, $\sum \limits_n V_n = 2 L$. A graph is $V$-regular if all nodes have the same valency, $V_n=V$, $\forall n$. A graph is regular if it is $V$-regular for some $V \in \mathbb{N}$.
\item \label{cond4} A graph is finite if both its node set and link set are finite. It is locally finite if the valency of each node is finite (the order of the graph can still be infinite).
\item \label{cond5} Two links intersect when they have a common node. A graph is connected if, for any pair of nodes $n,n'$, there is a sequence of links $\{\ell_i\}_{i=1}^r$ such that $n \in \ell_1$, $n' \in \ell_r$ and $\ell_i$ intersects $\ell_{i+1}$, for $i=1,\dots,r-1$. The graph distance $d(n,n')$ is the minimum number of links in any such a sequence connecting the nodes. Any graph decomposes into a family of maximal connected components.
\item \label{cond6} The complement of a graph $\Gamma$ is a graph $\Gamma^*$ with the same set of nodes as $\Gamma$ such that two nodes of $\Gamma^*$  are adjacent if and only if they are not adjacent in $\Gamma$. \item \label{cond7} Let $\tilde{N} \subset N$ be any subset of the set of nodes $N$ of a graph $\Gamma=\{N,L\}$. Then the induced subgraph $\tilde{\Gamma}=\{\tilde{N},\tilde{L}\}$ is the graph with node set $\tilde{N}$ and whose link set consists of all of links in $L$ that have both endpoints in $\tilde{N}$.
\item \label{cond8} The links, nodes, or both may be assigned specific values, labels, or colors, in which case the graph is called a labeled graph. A graph or directed graph together with a function that assigns a positive real number to each link is a network.
\end{itemize}

\subsection{Symmetries of graphs}

The symmetries of a graph are described by its automorphism group $\mathrm{Aut}(\Gamma)$. An automorphism $A \in \mathrm{Aut}(\Gamma)$ is a graph isomorphism, defined as a bijection $A: N(\Gamma)\rightarrow N(\Gamma)$ that preserves the adjacency matrix $\mathcal{A}_{nn'}$ of the graph \cite{biggs}. An automorphism is described by a permutation $\pi$ of the node set $N(\Gamma)$ such that any pair of nodes are adjacent if and only if their images under the permutation are also adjacent, $n \sim n' \Leftrightarrow \pi(n) \sim \pi(n')$.

Graph automorphisms are adjacency, valency and distance preserving. For oriented graphs, orientations must be preserved; for graphs with labeled links or nodes, labels must be preserved as well. Special families of symmetric graphs are defined by specific properties of their automorphism groups, such as~\cite{biggs}:
\begin{itemize}
\item \textit{Node-transitivity:} A graph is node-transitive if, for any pair of nodes $n,n' \in N(\Gamma)$, there is an automorphism $A$ such that $A(n) = n'$, i.e., if its automorphism group acts transitively on the nodes of the graph. 
\item \textit{Link-transitivity:} A graph is link-transitive if, for any pair of links $\ell,\ell' \in L(\Gamma)$, there is an automorphism $A$ such that $A(\ell)=\ell'$. Every link can be mapped by an automorphism into any other link.
\item \textit{Arc-transitivity:} A graph is arc-transitive if for any pair of oriented links $\ell,\ell' \in L(\Gamma)$, there is an automorphism $A$ such that $A(\ell)=\ell'$.
\item \textit{Distance-transitivity:} A graph is distance-transitive if, for any two pairs of nodes $(m,n)$ and $(m',n')$ with distances $d(m, n) = d(m',n')$, there is an automorphism $A$ such that $A(m)=m'$  and $A(n)=n'$.
\end{itemize}  

Node-transitivity is the weakest symmetry condition among those listed above or considered in this work. In a node-transitive graph, all individual nodes are equivalent, but this equivalence does not extend to regions formed by more than one node. The link-transitivity property is a stronger restriction, since all pairs of adjacent nodes are equivalent in a link-transitive graph. Distance-transitivity is even stricter as it involves the comparison of pairs of equidistant nodes that can be anywhere in the graph. Distance-transitive graphs are arc-transitive, and every arc-transitive graph is both link- and node-transitive. Families of highly symmetric \textit{homogeneous} graphs will be discussed in the next section.

\begin{figure}[htbp]
\includegraphics[scale=0.5]{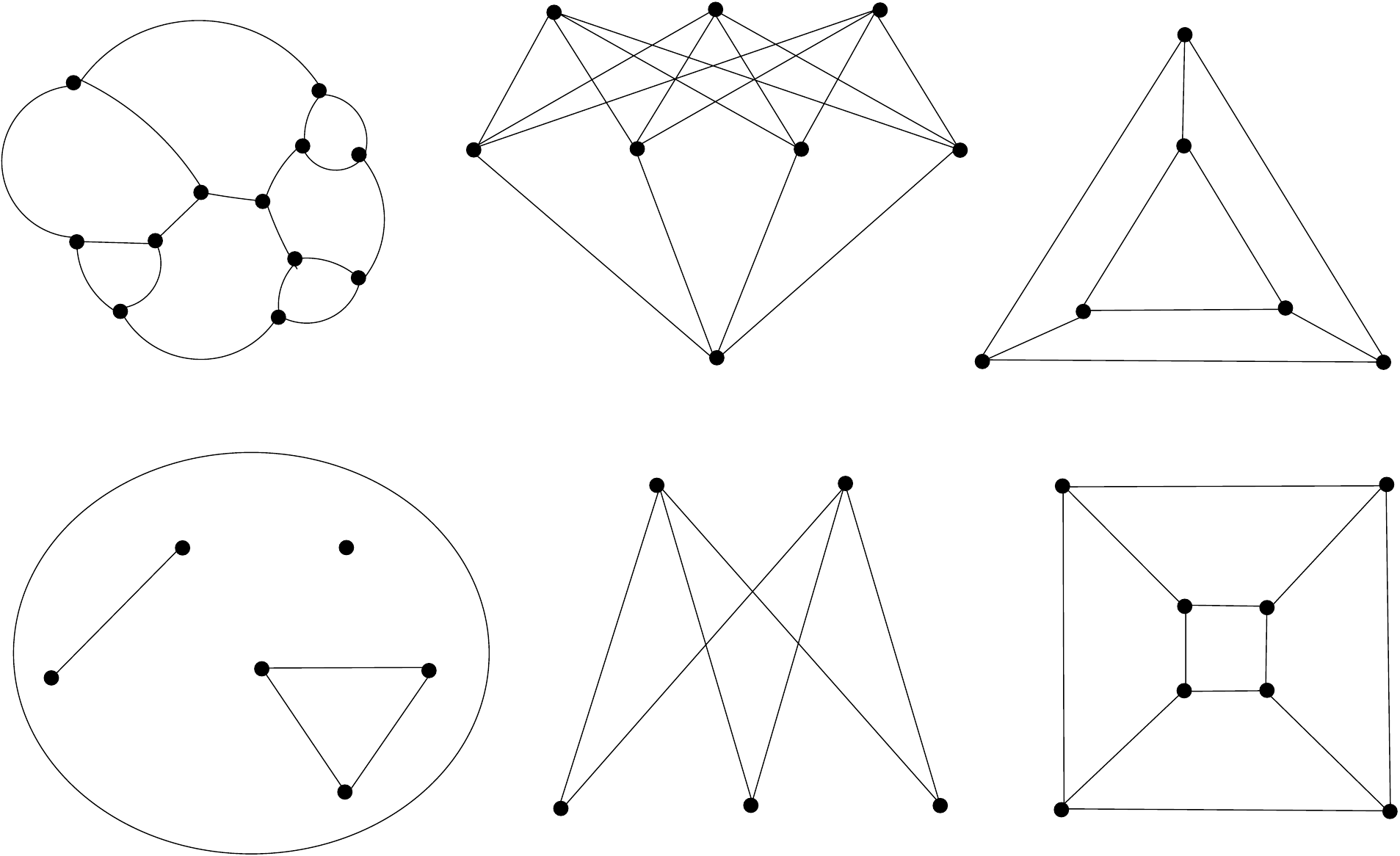}
\caption{\emph{Top left:} Frucht graph, a $3$-regular graph with $V=3$. It has no nontrivial symmetry: its automorphism group includes only the identity. \emph{Top middle:} The $4$-regular connected homogeneous graph with $V=8$.  \emph{Top right:} A node-transitive graph, the $3$-prism $\mathrm{Ci}_6[2,3]$, with $V=3$. \emph{Bottom left:} A cluster graph, the disjoint union (non-regular) of $K_1$, $K_2$ and $K_3$. \emph{Bottom middle:} A link-transitive graph, $K_{2,3}$. \emph{Bottom right:} An arc-transitive graph, the cubical graph $Q_3$ with $V=3$.}
\label{fig:graphs}
\end{figure}

Regularity is not a form of symmetry, in the sense of not being a property naturally associated with the automorphism group of the graph. The regularity of a graph can be checked directly from the adjacency matrix, in contrast with the properties listed above. A regular graph can have no nontrivial symmetry, for instance, as the Frucht graph, represented in Fig.~\ref{fig:graphs}. All node-transitive graphs are regular but not all regular graphs are node-transitive, and not all homogeneous graphs are regular (see Fig.~\ref{fig:graphs}).

\subsection{Maximally symmetric graphs: Homogeneous graphs}

A graph is called homogeneous if every isomorphism between two induced subgraphs extends to an automorphism of the graph~\cite{gardiner76}. In particular, a locally finite graph $\Gamma_{\mathrm{hom}}$ is called homogeneous if, for any isomorphic subgraphs $\Gamma_1, \Gamma_2$ with node sets $N_1(\Gamma_1), N_2(\Gamma_2) \subseteq N(\Gamma_{\mathrm{hom}})$ of the same order, $N_1=N_2 \leq N$, and for every isomorphism $\bar{A}: N_1(\Gamma_1) \rightarrow N_2(\Gamma_2)$, there exists an automorphism $A$ of $\Gamma_{\mathrm{hom}}$ such that $A|_{N_1(\Gamma_1)}= \bar{A}$ and $A^{-1} |_{N_2(\Gamma_2)} = \bar{A}^{-1}$. The classification of finite homogeneous graphs is known~\cite{gardiner76, ronse78}. Any such graph is isomorphic to one of the following graphs (see Appendix \ref{sec:graph-list} for the definition of the graphs involved in the classification below):
\begin{enumerate}
\item The cluster graph $m K_{N}$, a disjoint union of $m$ complete graphs $K_{N}$ with $N$ nodes;
\item Tur\'{a}n graphs, the complements $(m K_{N})^*$ of the cluster graphs;
\item The pentagon $C_5$;
\item The line graph $\mathrm{L}(K_{3,3})$ of the complete bipartite graph $K_{3,3}$, i.e., the $3 \times 3$ rook's graph.
\end{enumerate}
The only countably infinite homogeneous graphs $\Gamma^{\infty}_{\mathrm{hom}}$ are, up to isomorphisms~\cite{lachlan1980,cameron1913,gray}:
\begin{enumerate}
\item Disjoint unions of isomorphic complete graphs, where the size of each complete graph ($K_{\infty}$), the number of copies ($m \to \infty$), or both, are countably infinite, and complements of such unions;
\item The Rado graph;
\item The Henson graphs and their complements. 
\end{enumerate}
Note that there is no locally finite connected regular homogeneous graph that is infinite.

\begin{figure}[htbp]
 \includegraphics[scale=0.8]{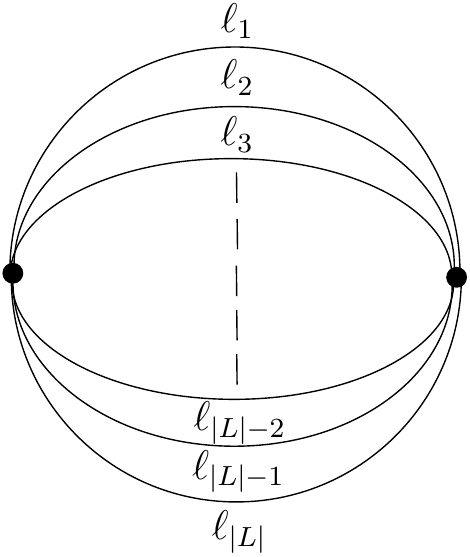}
\caption{The (undirected) dipole graph: a multi-graph without self-loops consisting of two nodes connected by $L$ links.}
\label{fig:dipolegraph}
\end{figure}

Homogeneous multi-graphs are not completely classified. The so-called dipole graph (see Fig.~\ref{fig:dipolegraph}), which consists of two nodes ($N = 2$) connected by an arbitrary finite number of links ($L$), is the simplest example of homogeneous multi-graph. We will mostly consider simple graphs in this work.

\begin{figure}[htbp]
 \includegraphics[scale=0.6]{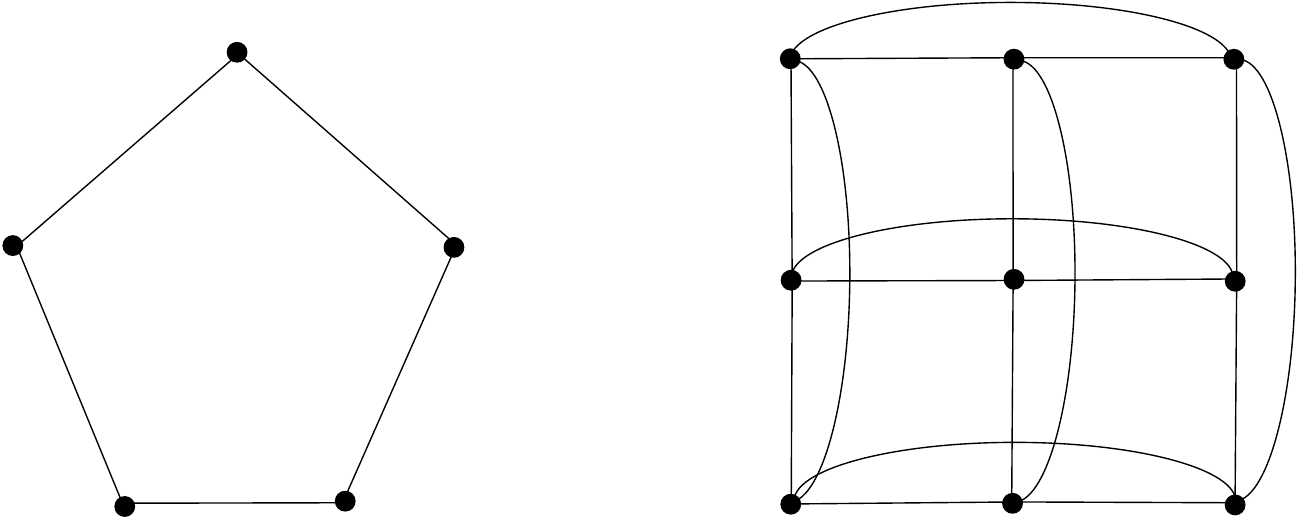}
\caption{\emph{Left:} The pentagon $C_5$. \emph{Right:} The Line graph $\mathrm{L}(K_{3,3})$ of $K_{3,3}$.}
\label{fig:pentagon-rook}
\end{figure}

A subclass of interest of the locally finite homogeneous graphs $\Gamma_{\mathrm{hom}}$ is obtained by restricting to connected and regular graphs. The classification of connected and regular locally finite homogeneous graphs, which we will shortly present, follows from the previous results. The pentagon $C_5$ and the rook's graph with fixed number of nodes and valency are in this class (see Fig.~\ref{fig:pentagon-rook}).  It also includes the complements of cluster graphs, $(m K_{N})^*$, which are examples of Tur\'{a}n graphs. Let us describe the relevant Tur\'{a}n graphs for this classification.

A Tur\'{a}n graph $\mathcal{T}(N', r)$ is a complete multi-partite graph obtained by first partitioning a set of $N'$ nodes into $r$ subsets, with sizes as close as possible, i.e., each two independent sets differing in size by at most one node, and then connecting two nodes by a link if and only if they belong to different subsets. It is regular if the number of subsets $r$ is a divisor of the number of nodes $N'$. The graphs $(m K_{N})^*$ are regular Tur\'an graphs, $(m K_{N})^*=\mathcal{T}(mN,m)$.

Any $V$-regular and connected locally finite homogeneous graph with $N$ nodes, except for $C_5$ and  $\mathrm{L}(K_{3,3})$, is the complement of a disjoint union of identical complete graphs $K_{N-V}$, 
\begin{align}
\bigg(\frac{N}{N-V} \, K_{N-V}\bigg)^* \,\,\, \longrightarrow \,\,\, \mathcal{T}\,\bigg(N,\frac{N}{N-V}\bigg) \, , 
\end{align}
which is the Tur\'{a}n graph with $N' \to N$ nodes and $r \to N/(N-V) \in \mathbb{N}\setminus\{0,1\}$ subsets.
\begin{figure}[htbp]
    \centering
    \subfloat[$(5 K_1)^* \to \mathcal{T}(5,5)$]{{\includegraphics[width=7cm]{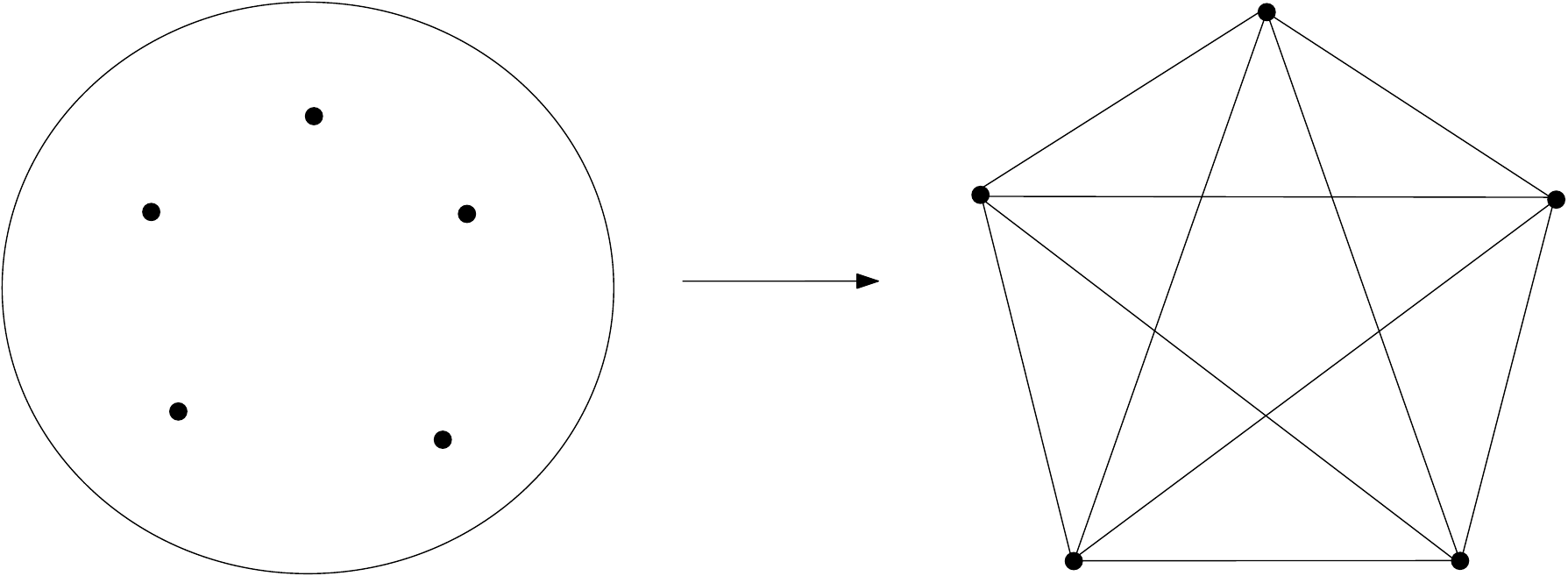} }}%
    \qquad
    \subfloat[$(3 K_2)^* \to \mathcal{T}(6,3)$]{{\includegraphics[width=7cm]{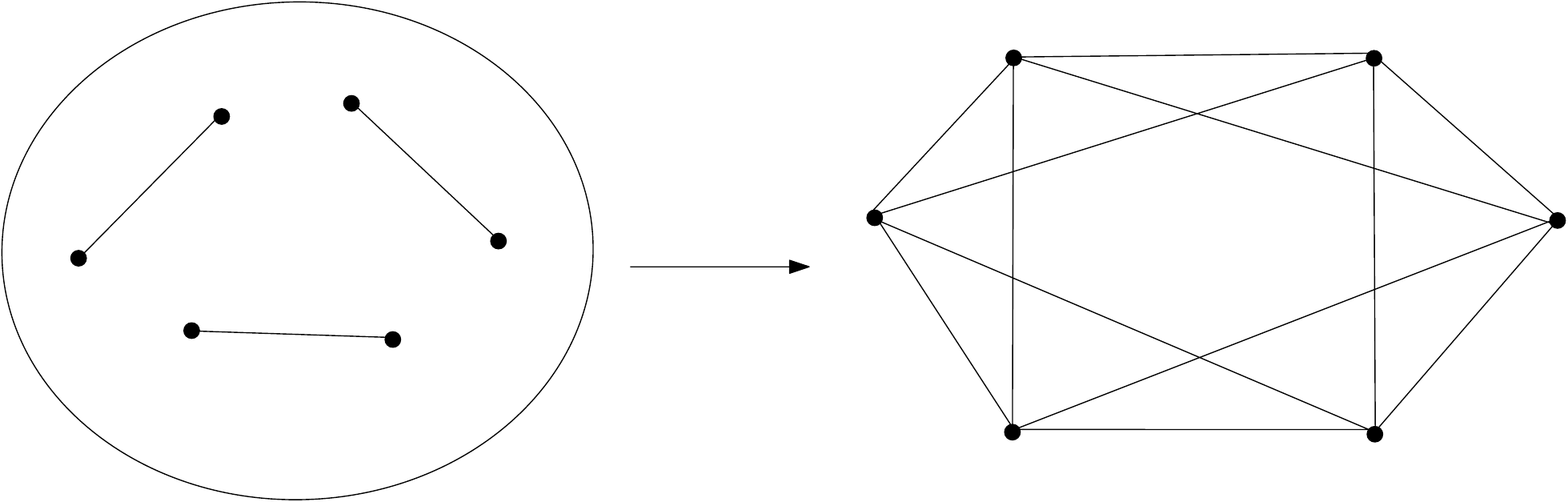} }}%
     \quad
     \subfloat[$(2 K_4)^* \to \mathcal{T}(8,2)$]{{\includegraphics[width=7cm]{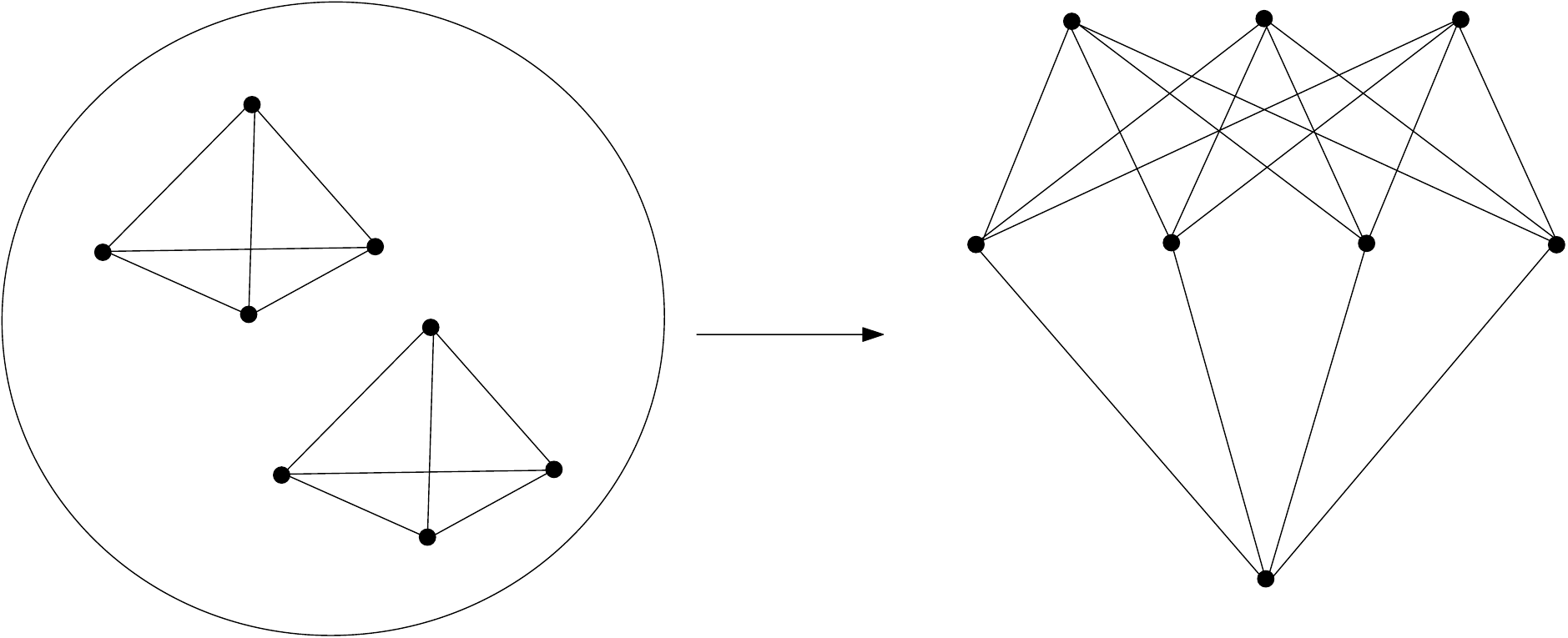} }}%
    \caption{The complete set of 4-regular Tur\'{a}n graphs, $\mathcal{T}(N,N/(N-4))$.}%
    \label{fig:4-regular-homogeneous}%
\end{figure}

The construction of the graphs $\mathcal{T}(N,N/(N-V))$ is illustrated in Fig.~\ref{fig:4-regular-homogeneous}, where all Tur\'an graphs with $V=4$ are represented. One starts with a collection of $N/(N-V)$ identical complete graphs $K_{N-V}$ with $N-V$ nodes. Next, the complement of $N/(N-V) K_{N-V}$ is taken by connecting nodes that are not adjacent in $N/(N-V) K_{N-V}$ and removing the links between adjacent nodes in each $K_{N-V}$.

The number of nodes $N$ in Tur\`an graphs of the form $\mathcal{T}(N,N/(N-V)$ is bounded by the valency $V$, which can be seen from the relation between $N$ and the number of independent subsets $r=N/(N-V)$,
\begin{align}
N= V \frac{r}{r -1}, \quad r >1 \quad \Longrightarrow \quad V+1 \leq N \leq 2 V \, .
\end{align}
This implies the existence of a finite number $\# \mathcal{T}_V$ of such graphs for a given finite valency $V$, equal to the number of positive divisors of $V$: 
\begin{align}
\# \mathcal{T}_V = (e_1+1) \cdots (e_k+1) \, , \quad \mathrm{if} \; V = \prod \limits_{i=1}^k \,p_i^{e_i} \, ,
\end{align}
where the $p_i$'s are prime numbers in the prime factorization of $V >1$, and $\# \mathcal{T}_1=1$, the only graph with $V=1$ being the graph $\mathcal{T}(2,2)=K_2$. For small valencies, the allowed number of nodes is given by:
\begin{align*}
V&=3: \quad N = {4,6} \, , \\
V&=4: \quad N = {5,6,8} \, , \\
V&=5: \quad N = {6,10} \, , \\
V&=6: \quad N = {7,8,9,12} \, .
\end{align*}
The number of links in regular Tur\`an graphs is also bounded by the valency,
\begin{equation}
\frac{1}{2} (V^2 +V) \leq L \leq 2 V^2 \, .
\end{equation}

The complete set of connected locally finite and $V$-regular homogeneous graphs is 
\begin{align}
\bigg\{ C_5,\mathrm{L}(K_{3,3}) ,\mathcal{T}\bigg(N,\frac{N}{N-V}\bigg) \, \big| \, V \geq 1\bigg\} \,,
\end{align}
the union of the pentagon, the rook's graph and the regular Tur\'{a}n graphs with $N$ nodes and $N/(N-V)$ subsets.

\subsection{Connected-homogeneous graphs}

The condition of homogeneity can be weakened in several ways. It is indeed of interest to consider weaker symmetry conditions for the introduction of discrete versions of cosmological spaces in loop quantum gravity. The classification of homogeneous graphs includes the pentagon $C_5$ but not the polygons $C_n$ with more than five nodes, for instance, while in a description of one-dimensional cosmological spaces it would be natural to include all polygons as viable graphs.

A graph is connected-homogeneous (C-homogeneous) if any isomorphism of connected induced subgraphs can be extended to the full graph. Any homogeneous graph is connected-homogeneous, but the converse is not true. The classification of C-homogeneous graphs is thus an enlargement of the list of homogeneous graphs discussed in the previous section. 

In a homogeneous graph, any two nodes $n,n'$ are equivalent: as the node $n$ and the node $n'$ are isomorphic as induced subgraphs, there is an automorphism of the graph relating them. This implies that any homogeneous graph is node-transitive. Now consider two pairs $(n_1,n'_1)$ and $(n_2,n'_2)$ of non-adjacent nodes. Then the induced subgraphs $\Gamma_1$ and $\Gamma_2$ associated with these sets are trivial graphs, since there is no link connecting $n_i$ to $n'_i$. As a result, they are isomorphic, and there is an automorphism of the graph that takes $n_1$ to $n_2$ and $n'_1$ to $n'_2$. Hence, any two pairs of nonadjacent nodes are equivalent in a homogeneous graph. As a result, the diameter of a homogeneous graph is at most two: if there were nodes such that $d(n,n')=3$, then this pair would not be equivalent to a pair of points with distance two, and the graph would not be homogeneous. This is the reason why polygons with more than five nodes are not homogeneous, having a diameter larger than two. The equivalence of all pairs of nonadjacent nodes is too restrictive a condition for the construction of cosmological spaces.

In a C-homogeneous graph, the argument cannot be reproduced. Since the graphs $\Gamma_i$ are not connected, the condition of C-homogeneity does not imply that the isomorphism between them can be extended to the full graph. The equivalence of all nodes and links still holds, however: C-homogeneous graphs are node-transitive and link-transitive. Moreover, local neighborhoods of any node can also be mapped isomorphically into local neighborhoods of any other point via the extended automorphism relating the nodes. Every C-homogeneous graph is also distance-transitive. The converse is also true for locally-finite graphs. Hence, for locally finite graphs, C-homogeneity is equivalent to distance-transitivity.

\begin{figure}[htbp]
 \includegraphics[scale=0.6]{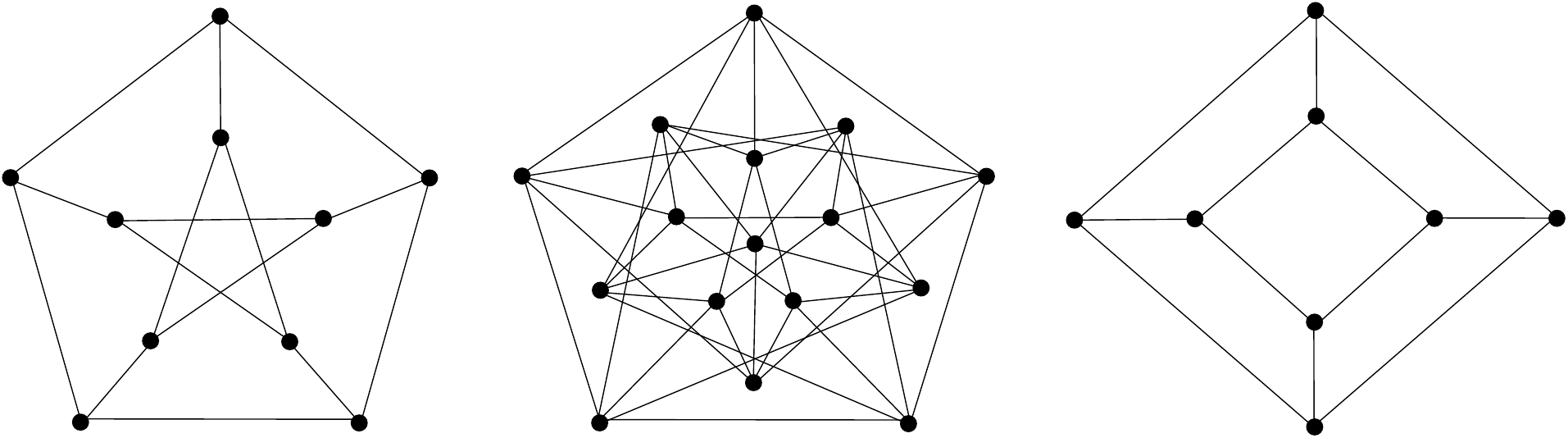}
\caption{\emph{Left:} Petersen graph $O_3$. \emph{Middle:} Clebsch graph $\Box_5$. \emph{Right:} Complement of a perfect matching $(2 \cdot K_{4})_{2}$, with $N=3$.}
\label{fig:CHgraphs}
\end{figure}

Special classes of C-homogeneous graphs have been classified, including the families of finite, locally-finite and countable C-homogeneous graphs. The classification of finite C-homogeneous graphs was obtained in \cite{gardiner78,enomoto}, and the classification of countable locally-finite C-homogeneous graphs was given in \cite{enomoto}. Countable C-homogeneous graphs were classified in \cite{gray}.

\begin{figure}[htbp]
 \includegraphics[scale=0.5]{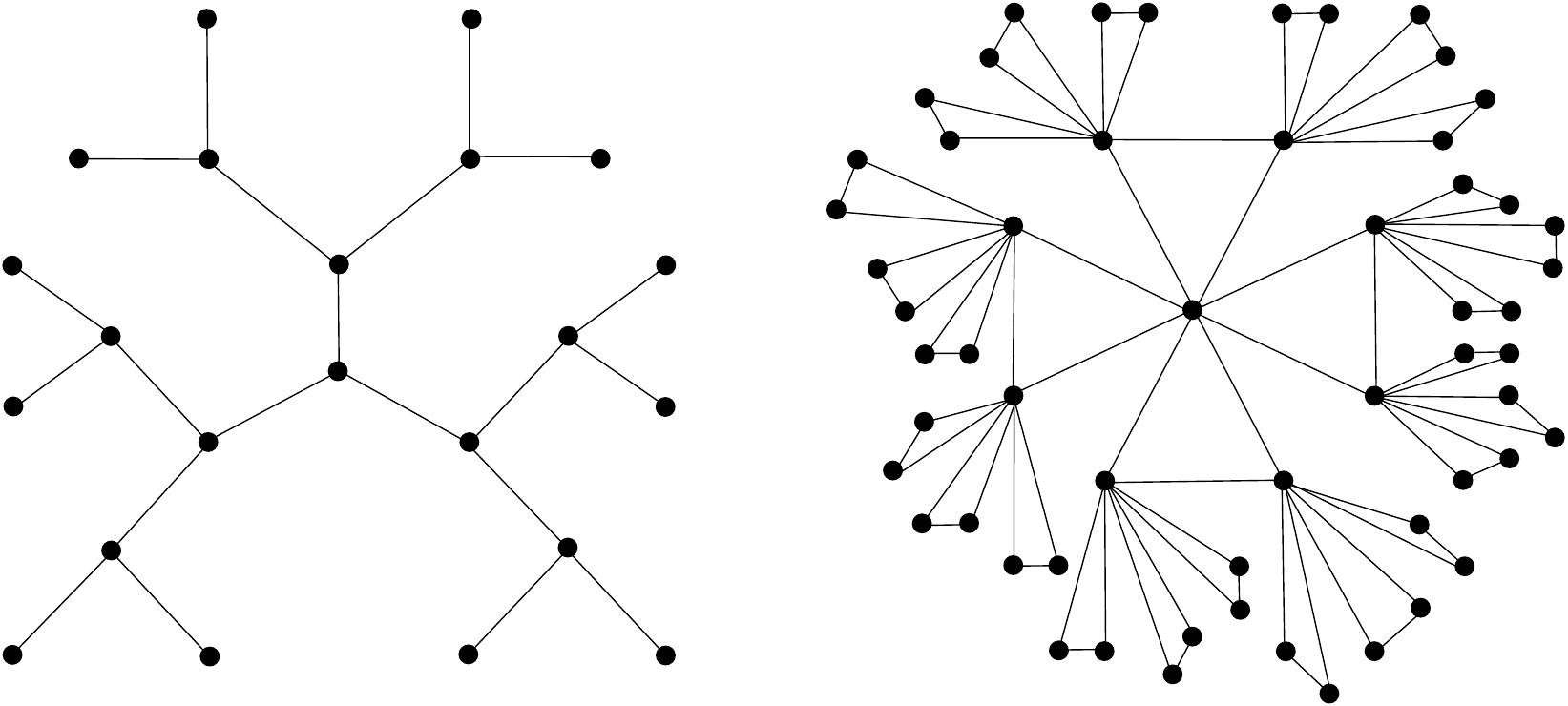}
\caption{ \emph{Left:} The infinite regular tree graph $X_{1,3}$ or Bethe lattice, with $V=3$. \emph{Right:} $X_{2,4}$, with $V=2$.}
\label{fig:infinitetreegraph}
\end{figure}

Any connected $V$-regular finite C-homogeneous graph (see Fig.~\ref{fig:CHgraphs}) is isomorphic to one of the following:
\begin{enumerate}
\item Regular Tur\'{a}n graphs: $\mathcal{T}(N,N/(N-V) \, (V \geq 1, \, V+1 \leq N \leq 2 V)$;
\item Complement of a perfect matching ($V \geq 2)$;
\item Line graph of a complete bipartite graph: $\mathrm{L}(K_{\frac{V}{2}+1,\frac{V}{2}+1}) \, (V \geq 4)$;
\item Cycle graphs: $C_{N} \, (V=2, N \geq 5)$;
\item Peterson graph: $O_3 \,  (V=3, N=10)$;
\item Clebsch graph: $\Box_5 \, (V=5, N=16)$.
\end{enumerate}
A countable C-homogeneous graph is the disjoint union of a countable number of isomorphic copies of a countable connected C-homogeneous graph. In particular, a finite C-homogeneous graph is a finite disjoint union of isomorphic copies of a finite C-homogeneous graph. The set of connected $V$-regular finite homogeneous graphs is a subset of that of connected countable, locally finite, and $V$-regular C-homogeneous graphs. Examples of connected, countably infinite, locally finite, and regular C-homogeneous graphs are represented in Fig.~\ref{fig:infinitetreegraph}.

\subsection{$k$-homogeneous and $k$-CH graphs}

The symmetry conditions of homogeneity and C-homogeneity are sufficiently restrictive to allow for the classification of graphs satisfying them. Weaker symmetry conditions are obtained by restricting the order of the isomorphic subgraphs required to be related by automorphisms of the graph, leading to the notion of $k$-homogeneity and $k$-CH graphs.

For any positive integer $k$, a graph is $k$-homogeneous if any isomorphism between induced subgraphs of order at most $k$ extends to an automorphism of the graph. The graph is homogeneous if it is $k$-homogeneous for all $k$. For each $k$, there are uncountably many countable $k$-homogeneous graphs that are not ($k + 1$)-homogeneous~\cite{cameron}. The locally finite $k$-homogeneous graphs have been classified for $2 \leq k \leq 4$ in \cite{buczak, cameron2, liebeck}.

A graph is $k$-C-homogeneous, or $k$-CH, if any isomorphism between connected induced subgraphs of order at most $k$ extends to an automorphism of the graph. It is C-homogeneous, or CH, if it is $k$-CH for all $k$. The 1-CH and 2-CH graphs are the node-transitive and arc-transitive graphs, respectively. The notions of $1$-homogeneous and $1$-CH graphs are equivalent. Any 2-CH graph is both node-transitive and link-transitive, as implied by arc-transitivity, as well as regular, by the node-transitivity.

The sets of C-homogeneous graphs and homogeneous graphs are subsets of the set of 2-CH graphs, since every connected C-homogeneous (and homogeneous) graph is arc-transitive. They match up to $N=9$ (with the exception of the cubic graph $Q_3$ for $N=8$ which is not homogeneous). Partial classifications of $2$-CH graphs were obtained by restricting the valency or the order of the graphs~\cite{wang,zhou,ghasemi}. For instance, any connected $2$-CH graph with valency $2$ is isomorphic to a cycle graph $C_{N}$.  Explicit lists of connected $2$-CH graph of a fixed valency up to a certain order have also been presented \cite{conder,potocnik}, which are extensions of the Foster census~\cite{foster}. 

Let us list some special families of $2$-CH graphs. All hypercube graphs $Q_n$ are connected $2$-CH graphs. The locally finite 2-CH graphs include the $n$-dimensional lattices $\mathcal{L}_{n}^{(\mathrm{V})}$and the dual graphs $\mathrm{Hc}(\{r,q,p\})$ of infinite regular honeycombs for three or higher dimensions with Schl\"{a}fli symbol $\{p,q,r\}$. The valency $V$ of the nodes of $\mathrm{Hc}(\{r,q,p\})$ is
 \begin{align}
 \label{eq:valency-Schlafli}
V = \frac{4 q}{2(p +q) - p q}\,, \quad p,q >2 \, .
 \end{align}
It must be finite for finite cells, which is only possible for the Platonic solids ($\{p,q\} = \{3,3\}$, $\{4,3\}$, $\{5,3\}$, $\{3,4\}$, $\{3,5\}$). The graphs formed by the vertices and edges of the Platonic solids are also $2$-CH.


\section{Homogeneity and isotropy in loop quantum gravity}
\label{sec:homogeneity-isotropy-lqg}

In this section, we discuss states in the Hilbert space of loop quantum gravity (LQG) that satisfy a discrete version of the conditions of homogeneity and isotropy. We first discuss maximally symmetric classical discrete geometries as a preparation for the quantum case. Then we introduce homogeneity and isotropy conditions for a quantum geometry through the selection of an adequate class of graphs and the restriction on the space of states and on the algebra of observables to objects that are invariant under the automorphism group of the graph.

\subsection{Maximally symmetric cell decompositions and dual graphs}
\label{sec:decompositions}

Let $\Delta$ be some polytopal decomposition of a $3$-manifold and $\Gamma$ its dual graph. A metric is introduced on the topological $3$-manifold $\Delta$ by the assignment of a specific geometry to each cell in the decomposition. If each topological polyhedron $\Delta^{(3)}_i \in \Delta$ is a flat polyhedron, for instance, we obtain a piecewise linear geometry.

The quantum geometry described by loop quantum gravity is the quantization of classical piecewise linear geometries called twisted geometries, in which glued faces always have the same total area but can have distinct shapes \cite{bianchi-polyhedra}. The metric can thus be discontinuous at the boundaries of the polyhedra in twisted geometries. We will discuss symmetries of classical piecewise linear geometries in this context before moving to the discussion of symmetries in the corresponding quantized geometries.

A polyhedron embedded in $\mathbb{R}^3$ is completely characterized by the area $\mathcal{A}_a$ of each face $a$ and the unit vector $\vec n_a$ normal to it. These quantities satisfy the closure relation
\be
\sum_{a=1}^V \mathcal{A}_a {\vec n}_a = {\vec 0 } \, ,
\label{eq:closure-relation}
\ee
which ensures that the data $(\mathcal{A}_a, \vec n_a)$ is uniquely associated with some polyhedron in $\mathbb{R}^3$, up to isometries of the ambient space \cite{minkowski,bianchi-polyhedra}. The space of shapes is characterized in this way by a simple condition on the variables $(\mathcal{A}_a, \vec n_a)$. This is not an invariant characterization of the intrinsic geometry of the polyhedron, however, as the unit normals $\vec n_a$ depend on how the polyhedron is embedded in $\mathbb{R}^3$.

The geometry of the polyhedron can be described in terms of invariant quantities, instead, as the areas $\mathcal{A}_a$ and angles $\theta_{ab}$ among the normals to the faces. Introducing the face vectors
\be
\label{eq:facevector}
{\vec E}_a=\mathcal{A}_a \vec n_a \, ,
\ee
we can write
\be
\label{eq:area}
\mathcal{A}_a = \sqrt{\vec E_a \cdot \vec E_a} 
\ee
and
\be
\cos \theta_{ab} = \frac{\vec E_a \cdot \vec E_b}{\sqrt{\vec E_a \cdot \vec E_a} \sqrt{\vec E_b \cdot \vec E_b}} \, .
\ee
All the invariant information required to reconstruct the intrinsic geometry of the polyhedron (its shape) is contained in the matrix
\be
\label{eq:Penrosemetric}
g_{ab} = \vec E_a \cdot \vec E_b \, ,
\ee
which provides a local description of the metric. After quantization, this matrix corresponds to the Penrose metric, which describes the local geometry of a quantum polyhedron in loop quantum gravity \cite{zakopane}.

We are interested in constructing piecewise linear approximations to homogeneous and isotropic $3$d geometries. Such maximally symmetric spaces admit special discretizations composed of simple blocks glued in a regular fashion. A simple example is the discretization of an Euclidean plane as a square lattice. Similarly, the two-dimensional sphere can be discretized as the boundary of a Platonic solid, and the two-dimensional hyperbolic space as a hyperbolic tessellation \cite{coxeter-whitrow}. These examples share the property that their dual graphs are $2$-CH graphs. Regular discretizations with dual $2$-CH graphs also exist in higher dimensions for maximally symmetric spaces of positive, zero and negative curvature \cite{coxeter-whitrow}, thus providing a natural setup for the discretization of generic cosmological spaces.

In three dimensions, the building blocks of a potytopal decomposition are polyhedra. Some examples of regular decompositions of homogeneous and isotropic 3d spaces with dual $2$-CH graphs are listed below \cite{coxeter,coxeter-whitrow}.
\begin{itemize}
\item $\mathds{R}^3$: Graphs dual to regular decompositions of $\mathds{R}^3$ are given by infinite 3d lattices $\mathcal{L}^{(\mathrm{V})}_{3}$ with valence $V \geq 4$. Each $V$-valent node is dual to a convex polyhedron with $V$ faces and each link incident to the node is dual to a polygonal face. The canonical example is the infinite cubic lattice or cubulation: $\mathcal{L}^{(\mathrm{6})}_{3}$. 
\item $\mathds{H}^3$: Graphs dual to regular decompositions of the hyperbolic space $\mathds{H}^3$ are given by graphs $\mathrm{Hc}(\{r,q,p\})$ with valency $V \geq 4$ (given by~\eqref{eq:valency-Schlafli}), dual to the infinite hyperbolic regular 3d honeycombs with Schl\"{a}fli symbol $\{p,q,r\}$. The values of $p$, $q$ and $r$ satisfy the condition 
\[\mu = \sin\left(\frac{\pi}{p}\right) \sin\left(\frac{\pi}{r}\right) - \cos\left(\frac{\pi}{q}\right) < 0 \, , \qquad p,r,q >2 \, .
\]
Examples are given by the regular hyperbolic honeycombs with Schl\"{a}fli symbols $\{3,5,3\}, \{4,3,5\}, \{5,3,4\},\{5,3,5\}$.
\item $S^3$: Examples of graphs dual to regular decompositions of $S^3$ are provided by finite connected C-homogeneous graphs with valencies $V \geq 4$, which include: $\Box_5$, $\mathcal{T}\big(N,\frac{N}{N-V}\big)$, $(2 \cdot K_{V+1})_{V-1}$, $\mathrm{L}(K_{\frac{V}{2}+1,\frac{V}{2}+1})$.
\end{itemize}

Piecewise linear geometries composed of identical polytopes glued along a $2$-CH dual graph thus form a natural class of discrete approximations of spatial sections of cosmological spacetimes. It is convenient to require, in addition, that the graph automorphisms are associated with isometries that map the polyhedra onto each other isometrically\footnote{This prevents, for instance, rectangles to be employed in a decomposition of the Euclidean plane based on a square lattice. A rotation of $\pi/2$ is an isometry of the space and is associated with a graph automorphism, but does not map the rectangular building blocks isometrically onto each other.}. This ensures that the group of isomorphisms preserving the combinatorial structure of a $2$-CH graph mirrors the key properties of the group of isometries of maximally symmetric continuous Riemannian geometries, as we wish to discuss now. Analogous conditions will later be imposed on states of the geometry in loop quantum gravity for the description of cosmological spaces.

Let $\Delta$ be a regular decomposition of a cosmological space and $\Gamma$ its dual graph. We assume that any automorphism $A$ of the dual graph is associated with an isometry of the space that maps the polyhedra isometrically among themselves. Since the graph $\Gamma$ is $2$-CH, it must also be $1$-CH. This means that the automorphism group acts transitively on the nodes. As a result, the polyhedra represented by the nodes of the dual graph are all equivalent, in the sense that there exists an isometry relating them, just as for points in a homogeneous space. In short, the $1$-CH property corresponds to the homogeneity of the decomposition. In addition, the $2$-CH property corresponds to the isotropy of the decomposition. The automorphism group of a $2$-CH graph acts transitively on its arcs; in particular, any two links emanating from a given node can be related by an automorphism of the graph that preserves the source node. This means that all directions from a give node along the dual graph of the decomposition are equivalent in a $2$-CH graph, in a manner analogous to the equivalence of all directions from a given point in an isotropic space. 

Any $k$-CH graph with $k>2$ is also a $2$-CH graph and thus satisfies the properties discussed above. In addition, regions of the graph larger than individual nodes and arcs are also all equivalent along the graph. For instance, in a $3$-CH graph, any pair of induced subgraphs with three nodes can be related by an automorphism of the graph. As each such subgraph naturally describes a wedge, that is, a pair of links emanating from a common node, then all wedges are equivalent in a $3$-CH graph. For a graph with valency $V=4$, the $3$-CH property implies that the nodes represent regular tetraheda: the fact that any two wedges can be mapped into each other means that all dihedral angles in the decomposition are equal.

We will restrict in what follows to connected graphs, so that the associated geometries are connected, and to locally finite graphs, in order that the polyhedra of the decomposition have a finite number of faces. As all nodes must be equivalent, they must have the same valency $V$. We denote a generic connected locally finite $1$-CH graph by the symbol $\Gamma_H$, or by $\Gamma^{(\mathrm{\scriptsize V})}_{H}$ when of interest to explicitly specify the valence $V$ of its nodes.  Similarly, we denote a generic connected locally finite $2$-CH graph with valency $V$ by the symbol $\Gamma_C$, or by $\Gamma^{(\mathrm{\scriptsize V})}_C$. For the description of 3d spaces, we further restrict to $V \geq 4$. The classes of homogeneous, C-homogeneous and higher $k$-CH graphs constitute special families of highly symmetric $2$-CH graphs that can be explored for the construction of the simplest concrete examples of cosmological discrete spaces.

\subsection{Space of states and observables in loop quantum gravity}
\label{sec:lqg-rev}

Let us briefly review the basics of loop quantum gravity \cite{rovellibook, ashtekarjerry, thiemannbook}. The kinematical Hilbert space $\mathcal{H}_{\mathrm{kin}}$ of the theory is defined as (see \cite{zakopane,rovellicombi})\footnote{We adopt the combinatorial definition of the space of states \cite{zakopane,rovellicombi}. Alternatively, the physical states can be defined over equivalence class of embedded graphs $\gamma$ on a fixed three-dimensional manifold $\Sigma$ under extended diffeomorphisms $\phi \in \mathrm{Diff}^*(\Sigma)$ \cite{fairbairn,zakopane}. Graphs in distinct knot classes then define distinct Hilbert subspaces of states of the geometry. In another approach, motivated by category theory, the group of diffeomorphisms of $\Sigma$ is extended to the automorphism group of the path groupoid of $\Sigma$ \cite{bahr}. In this case, the states of the geometry do not depend on how the graphs are embedded in $\Sigma$, leading to a space of states that is equivalent to that obtained through the combinatorial definition.}
\begin{align}
\label{eq:kinematical}
\mathcal{H}_{\mathrm{kin}} = \bigoplus_{\Gamma} \, \mathcal{H}_{\Gamma} =  \bigoplus_{\Gamma, \{j_\ell\}} \, \mathcal{H}_{\Gamma, \{j_\ell\}} \,.
\end{align}
The first sum runs over all oriented graphs $\Gamma$, and the second sum runs over all oriented graphs $\Gamma$ and spins $j_\ell=n/2$, with $n = 1,2, \dots$, where the spins are assigned to the unoriented links of the graph, so that $j_\ell=j_{\ell^{-1}}$. The space $\mathcal{H}_{\Gamma}$ is isomorphic to the Hilbert space of gauge-invariant states $\Psi_{\Gamma}(h_\ell) \in L^2_{\Gamma}[\mathrm{SU(2)}^{L}/SU(2)^N]$ over the graph $\Gamma$ that satisfy the invariance condition:
\begin{align}
\Psi_{\Gamma}(h_\ell) = \Psi_{\Gamma}(U_{s(\ell)} \, h_\ell \, U^{-1}_{t(\ell)})\,, \quad \forall U_{s(\ell)}, U_{t(\ell)} \in \mathrm{SU(2)} \, ,
\end{align}
where the labels $s(\ell)$ and $t(\ell)$ describe the source and target nodes of the link $\ell$.

The Hilbert space $\mathcal{H}_{\Gamma}$ decomposes into a direct sum over subspaces $\mathcal{H}_{\Gamma, \{j_\ell\}}$ with fixed configurations $\{j_\ell\}$ of nonzero spins. Each subspace $\mathcal{H}_{\Gamma, \{j_\ell\}}$ has the tensor product structure
\be
\mathcal{H}_{\Gamma, \{j_\ell\}} = \bigotimes_{n=1}^{N}  \, \mathcal{H}_n\,,
\label{eq:node-spaces-product}
\ee
where $\mathcal{H}_n$ is the space of the $\mathrm{SU(2)}$ intertwiner states $\ket{i_{k_n}}$ at the individual nodes $n$,
\be
\mathcal{H}_n = \mathrm{Inv}_{\mathrm{SU(2)}} \left[ \left( \bigotimes_{s(\ell)= n} \, \mathcal{V}_{j_\ell} \right) \otimes \left( \bigotimes_{t(\ell)=n} \mathcal{V}^*_{j_\ell}\right) \right]
\ee
The tensor product at $n$ involves a representation space $\mathcal{V}_{j_\ell}$ of spin $j_\ell$ of $SU(2)$ if $\ell$ points outwards from the node, or its dual $\mathcal{V}^*_{j_\ell}$, if the link is oriented towards the node.

A linear span of $\mathcal{H}_{\Gamma}$ is provided by the family of spin network states
\begin{align}
\ket{\Gamma, j_\ell,i_n} = \bigotimes_n \ket{i_n} \,,
\end{align}
labelled by the spins $j_\ell$ and an orthonormal basis $\ket{i_n}$ of $\mathcal{H}_n$ attached to the nodes of the graph. More explicitly, under the isomorphism $\mathcal{H}_\Gamma \cong L^2_{\Gamma}[\mathrm{SU(2)}^{L}/SU(2)^N]$, the wavefunction of a spin network state $\ket{\Gamma, j_\ell, i_n}$ is
\be
\psi_{\Gamma, j_\ell, i_n}(h_\ell) = \sum_{m_{na}=-j_{na}}^{j_{na}} \left(\prod_n [i_n]^{\,\, m_{n1} \cdots m_{n \mu}}{}_{\, m_{n (\mu+1)} \cdots } \right) \left( \prod_\ell \left[D^{j_\ell}(h_\ell)\right]^{m_{t(\ell)}}{}_{m_{s(\ell)}} \right) \, ,
\label{eq:spin-network-def}
\ee
where $D^{j_\ell}$ and $i_n$ are Wigner matrices and intertwiner tensors in the magnetic number basis, respectively. The set of indices $a=1,\dots,V$ is an ordering of the links $\ell=na$ at the $V$-valent node $n$, and the magnetic numbers $m_{s(\ell)}, m_{t(\ell)}$ are at the source and target nodes of $\ell$. The number $\mu$ of upper indices in $i_n$ corresponds to the number of links pointing outwards from the node; the lower indices correspond to links pointing towards the node.

Although a choice of orientation for the links is employed in the definition of the states $\Psi_\Gamma$, it does not carry any physical meaning. In fact, orientation reversal operations are introduced together with an equivalence relation that factors out the arbitrary choice of orientations. Let $R_\ell \Gamma$ be the graph obtained from $\Gamma$ by reversing the orientation of a link $\ell$. Denote by $\ell^{-1}$ the link $\ell$ with the reverse orientation, and put $h_{\ell^{-1}}=h_\ell^{-1}$. A state in $\mathcal{H}_{\Gamma}$ is then mapped into a state in $\mathcal{H}_{R_\ell \Gamma}$ by the isometry:
\begin{align}
\Psi(h_1, \dots, h_\ell, \dots, h_L) \mapsto (R_\ell \Psi)(h_1, \dots, h_{\ell^{-1}}, \dots, h_L) = \Psi(h_1, \dots, h_\ell, \dots, h_L) \, .
\end{align}
An equivalence relation is then introduced, $\Psi_\Gamma \sim R_\ell \Psi_\Gamma, \, \forall R_\ell$, in the direct sum $\mathcal{H}_{\bar{\Gamma}} = \bigoplus \mathcal{H}_{\Gamma}$, where the sum runs over all oriented graphs $\Gamma$ on the unoriented graph $\bar{\Gamma}$. In this way, the same amplitude is assigned for a given configuration of holonomies regardless of the orientation of the links, under the identification $h_{\ell^{-1}}=h_\ell^{-1}$.  A representative of each equivalence class can be specified by fixing an orientation for all links arbitrarily and letting $h_\ell$ refer to this orientation.

A holonomy operator $h_\ell$ and a flux operator $\vec E_{\ell}$ are associated with each link $\ell$ equipped with the orientation specified by $\Gamma$. These operators generate the holonomy-flux algebra $\mathfrak{A}$ of observables of loop quantum gravity. Their commutators are:
\begin{align}
\label{eq:holonomyfluxalgebra}
[h_\ell, h_{\ell'} ] = 0\,, \quad [E^i_\ell, E^j_{\ell'} ] = i \mathfrak{a}_{0} \delta_{\ell \ell'} \, \epsilon^{ij}_{\,\,\,\, k}  \, E^k_{\ell} \,, \quad [h_\ell, E^i_{\ell'} ] = -\textstyle{\frac{1}{2}} \, \mathfrak{a}_{0} \, \delta_{\ell \ell'} \, h_\ell \, \sigma^i \,,
\end{align}
where $\mathfrak{a}_{0}=8\pi G\hbar \gamma$ is a constant with units of area, $\sigma^i$ are Pauli matrices and $\epsilon^{ij}_{\,\,\,\, k}$ is the Levi-Civita symbol with $\epsilon_{123}=1$. Operators associated with distinct links commute. The holonomy operators act as multiplication operations in the holonomy representation. The flux operator $\vec {E}_\ell = \mathfrak{a}_{0} \vec J_\ell$ is proportional to the left-invariant vector field $\vec J_\ell$ that acts in the holonomy representation as a derivative of the wavefunction $\Psi(h_1,\dots,h_L)$ with respect to the holonomy $h_\ell$.

The flux operators are the quantization of the face vectors introduced in Eq.~\eqref{eq:facevector}. If $n=s(\ell)$, then $\vec{E}_\ell=\vec{E}_{na}$ is precisely the operator corresponding to the classical face vector at the face corresponding to the link $a$ at the node $n$. Each intertwiner space $\mathcal{H}_n$ corresponds to a quantum polyhedron associated with the node $n$ \cite{bianchi-polyhedra}. The intertwiner states $\ket{i_n}$ satisfy the Gauss constraint 
\begin{equation}
\sum_{a=1}^{V} \vec E_{na} \, \ket{i_n} =0 \, ,
\end{equation} 
which is the quantized version of the closure relation \eqref{eq:closure-relation}. The flux operators describe the intrinsic geometry of the quantum polyhedron at the node $n$ through the promotion of the matrix $g_{ab}$ defined in Eq.~\eqref{eq:Penrosemetric} to the Penrose metric (or shape) operator
\begin{equation}
\label{eq:shapeoperator}
\hat{g}_{ab}(n) = \vec E_{na} \cdot \vec E_{nb} \, .
\end{equation}
The area and dihedral angle operators are then given by the formulas \eqref{eq:facevector} and \eqref{eq:area}, now written in terms of flux operators $\vec E_{na}$,
\begin{align*}
\hat{\mathcal{A}}_{na}&=\sqrt{\hat{g}_{aa}(n)} \, , \\
\cos \hat{\theta}_{ab}(n) &= \frac{\hat{g}_{ab}(n)}{\sqrt{\hat{g}_{aa}(n)} \sqrt{\hat{g}_{bb}(n)}} \, , \quad \text{for }a\neq b \,.
\end{align*}
The intertwiner states $\ket{i_n}$ are eigenstates of the area operator,
\begin{equation}
\hat{\mathcal{A}}_{na} \, \ket{i_n} = \mathfrak{a}_{0}\,\sqrt{j_{na}(j_{na}+1)} \, \ket{i_n} \,.
\end{equation}
Heisenberg uncertainty relations for the quantum geometry hold as a result of the non-commutativity of different components of the shape operator $\hat{g}_{ab}(n)$,
\begin{align}
\label{eq:algebraofobservables}
[\hat{g}_{ab}(n), \hat{g}_{ac}(n)] &= i \,\mathfrak{a}_{0} \;\vec E_{na} \!\cdot \!(\vec E_{nb} \times \vec E_{nc}) \nonumber \, , \\ 
[\hat{g}_{ab}(n), \hat{g}_{aa}(n)] &= [\hat{g}_{aa}(n), \hat{g}_{bb}(n)] = 0 \,, \quad a < b < c \leq V \,.
\end{align}
The dispersions $\Delta \hat{g}_{ab}(n)$ in the quantum shape of the polyhedron satisfy the inequality
\be
\Delta \hat{g}_{ab}(n) \, \Delta \hat{g}_{ac}(n) \geq \frac{\mathfrak{a}_{0}}{2}  \, \bigg| \langle i_{n}| \vec E_{na} \cdot (\vec E_{nb} \times \vec E_{nc} ) |i_{n}\rangle \bigg| \, .
\label{eq:uncertainty-relations}
\ee 
We will omit the hats over operators of the quantum geometry from now on.

\subsection{Automorphism invariant states and observables}
\label{sec:graphautomorphisms}

\paragraph{Automorphism-invariant states.}

If the automorphism group $\mathrm{Aut}(\Gamma)$ of the graph $\Gamma$ is nontrivial, then the physical states of the geometry must be invariant under the graph automorphisms \cite{rovellicombi}. This condition is analogous, in the combinatorial definition of the space of states, to the invariance under diffeomorphisms in the construction of the space of states based on graphs embedded on some fixed differentiable manifold \cite{arrighi,colafranceschi}. It ensures that the definition of the states depends only on the combinatorial structure of the graph, and not on the choice of a particular representation of the graph. States $\ket{\Psi_{\Gamma}}$ and $\ket{\Psi'_{\Gamma}}$ in $\mathcal{H}_{\Gamma}$ are said to be equivalent under graph automorphisms if they are related by an automorphism of $\Gamma$.

The action of an automorphism $A:\Gamma \to \Gamma$ on $\mathcal{H}_\Gamma$ is given by the unitary operator defined by
\begin{align}
 \label{eq:action-automorphism}
\Psi(h_\ell) \mapsto (U_A \Psi)(h_\ell) = \Psi'(h_\ell) = \Psi(h_{A(\ell)}) \, ,
\end{align}
or, inserting the arguments explicitly,
\[
\Psi'(h_{\ell_1},\dots,h_{h_{\ell_L}}) = \Psi(h_{A(\ell_1)},\dots,h_{A(\ell_L)}) \, .
\]
Under this action, the wavefunction is simply carried along by the automorphism, and the argument of the wavefunction is inverted for links whose orientation is flipped by the automorphism. We denote by $\mathcal{K}_{\Gamma} \subset \mathcal{H}_{\Gamma}$ the subspace formed by states invariant under the automorphism group $\mathrm{Aut }(\Gamma)$ of the graph $\Gamma$, and by $P_{A}: \mathcal{H}_{\Gamma} \to \mathcal{H}_{\Gamma}$ the projector to such invariant subspace,
\be
\label{eq:invariantsubspace}
P_A \, \mathcal{H}_{\Gamma} = \mathcal{K}_{\Gamma} \, .
\ee

For graphs with a finite automorphism group, the kinematical Hilbert space $\mathcal{K}_{\Gamma}$ can be obtained from $\mathcal{H}_{\Gamma}$ by a procedure of group averaging under the action of $\mathrm{Aut}(\Gamma)$.  The projector is given for such graphs by
\be \label{eq:diff-group-averaging}
P_A = \frac{1}{|\mathrm{Aut}(\Gamma)|} \,\, \sum_{A \,  \in  \, \mathrm{Aut}(\Gamma)} U_A \, .
\ee
A state described by a density matrix $\rho^{\mathrm{inv}}_\Gamma$ in $\mathcal{K}_{\Gamma}$ is invariant under the automorphism group $ \mathrm{Aut}(\Gamma)$ when
\begin{align}
\label{eq:invariantstate}
U_A \,\, \rho^{\mathrm{inv}}_\Gamma \,\, U_A^{-1} = \rho^{\mathrm{inv}}_\Gamma \, , \qquad \forall U_A \in \mathrm{Aut }(\Gamma) \, .
\end{align}
For graphs with an infinite automorphism group, the space $\mathcal{K}_{\Gamma}$ is still defined as the space of states invariant under the action of the symmetries, but it cannot be constructed by a procedure of group averaging, since $|\mathrm{Aut}(\Gamma)| = \infty$ in this case.

The action of automorphisms on spin network states is obtained by the direct application of Eq.~\eqref{eq:action-automorphism}. The transformation moves the spins and intertwiners around according to its action on the graph. Let us describe this action more explicitly. 

Consider a link $\ell$ for which $A(\ell)=\bar{\ell}^{-1}$, where $\bar{\ell}$ is a link of the oriented graph $\Gamma$. In this case, the orientation of the transformed link is the opposite of that originally fixed in the graph. Let $m=s(\ell)$ and $n=t(\ell)$ be the source and target nodes of the link $\ell$. Denote their images under the automorphism $A$ by $\bar{m}=A(m)$, $\bar{n}=A(n)$.  The spin network state $\ket{\Gamma, j_\ell, i_n}$ depends on $h_\ell$ through
\be
[i_m]^{\alpha \cdots} \, [D^{j_\ell}(h_\ell)]^\beta{}_\alpha \, [i_n]_{\beta \cdots}
\ee
After the application of the automorphism, the new state depends on the holonomy $h_{\bar{\ell}}$ through
\be
[i_m]^{\alpha \cdots} \, [D^{j_\ell} (h_{\bar{\ell}}^{-1})]^\beta{}_\alpha \, [i_n]_{\beta \cdots} \, .
\ee

We raise and lower indices with the isomorphisms:
\be \label{eq:epsilon-isomorphism}
v^m = \epsilon^{(j)mn}v_n \, , \quad v_m = v^n \epsilon^{(j)}{}_{nm} \, ,
\ee
where
\be
\epsilon^{(j)mn} = \epsilon^{(j)}{}_{mn} = (-1)^{j-m} \delta_{m,-n} \, .
\ee
In addition, the Wigner matrix satisfies the identity:
\be
[D^{j}(g^{-1})]^m{}_n = \epsilon^{(j)mm'} \epsilon^{(j)}{}_{nn'} [D^{j}(g)]^{n'}{}_{m'} \, .
\ee
It follows that the transformed state satisfies:
\begin{align}
\label{eq:inversecontraction}
[i_m]^{\alpha \cdots} \, [D^{j_{\ell}}(h_{\bar{\ell}}^{-1})]^\beta{}_\alpha \, [i_n]_{\beta \cdots} 
= (-1)^{2 j_\ell}[i_{m}]_\alpha{}^{\cdots} \, [D^{j_{\ell}}(h_{\bar{\ell}})]^\alpha{}_\beta  \, [i_{n}]^\beta{}_{\cdots} \, .
\end{align}

If, on the other hand, the orientation of a transformed link agrees with the original orientation, so that $A(\ell)=\bar{\ell}$ for some $\bar{\ell}$, there is no need to raise or lower indices in the intertwiners. We conclude that the transformed state is a new spin-network state
\be
\label{eq:autoaction}
U_A \, \ket{\Gamma, \{j_\ell\}, \{i_n\}} = s_{A,j_\ell} \ket{ \Gamma, \{ j'_\ell \}, \{i'_n\}}  \, ,
\ee
with the configurations of spins and intertwiners carried by the automorphism, 
\[
j'_\ell = j_{A^{-1}(\ell)} \, , \qquad i'_n = i_{A^{-1}(n)} \, ,
\]
and intertwiner indices raised and lowered when necessary with the isomorphism \eqref{eq:epsilon-isomorphism}. The factor $s_{A,j_\ell}=(-1)^R$ is a possible sign flip, with $R$ corresponding to the number of links with semi-integer spins whose image under $A$ has an orientation that disagrees with that of the graph. The action of an automorphism $A$ defines permutations of the spins and intertwiners, which we will represent as
\begin{align*}
( j_\ell,i_n ) \mapsto A( j_\ell,i_n ) &= (A {j_\ell,A i_n}) \\
	&=(j'_\ell,i'_n) \\
	&=(j_{A^{-1}(\ell)}, i_{A^{-1}(n)}) \, .
\end{align*}

A generic state $\ket{\Psi^{\mathrm{inv}}_{\Gamma}}$ in $\mathcal{K}_\Gamma$ is the projection of a superposition of spin-network states,
\[
\ket{\Psi^{\mathrm{inv}}_{\Gamma}} = P_A \ket{\Psi_{\Gamma}} \, , \quad \ket{\Psi_{\Gamma}} = \sum \limits_{j_\ell, i_n} c_{j_{\ell}, i_n} \ket{\Gamma, \{j_{\ell}\}, \{i_n}\} \, .
\]
Distinct spin-network states can have the same image under the projection $P_A$. We introduce an equivalence relation $\ket{\Gamma, \{j_{\ell}\}, \{ i_n\}} \sim \ket{\Gamma, \{j'_{\ell}\}, \{i'_n \}}$ for states with the same image, $P_A\ket{\Gamma, \{j_{\ell}\}, \{i_n\}} = P_A\ket{\Gamma, \{j'_{\ell}\}, \{i'_n\}}$. The equivalence classes of spin-network states under such a relation  form a basis of $\mathcal{K}_\Gamma$.

A special class of states in $\mathcal{K}_\Gamma$ is given by the completely symmetric states $\ket{\Psi_\Gamma}$ invariant under the permutation group $\mathcal{S}_N$ acting on the nodes of the graph. When the graph $\Gamma$ is a complete graph $K_N$ with $N$ nodes, the completely symmetric states span the full space $\mathcal{K}_{K_N}$.

\paragraph{Automorphism-invariant observables.}

An operator $\mathcal{O}_\mathrm{inv}$ acting on the Hilbert space $\mathcal{H}_\Gamma$ is invariant under automorphisms if $\mathcal{O}_\mathrm{inv} = U_A \mathcal{O}_\mathrm{inv} U_A^{-1}$, $\forall U_A \in \mathrm{Aut}(\Gamma)$. Given any observable $\mathcal{O}$, not necessarily automorphism-invariant, we can construct an observable $\mathcal{O}_{\mathrm{inv}}$ that is automorphism-invariant through
\begin{equation}
\mathcal{O}_{\mathrm{inv}} = \frac{1}{|\mathrm{Aut }(\Gamma)|} \,\, \sum_A U_A \, \mathcal{O} \, U_A^{-1} \, .
\label{eq:O-inv}
\end{equation}
The subspace of automorphism-invariant states $\mathcal{K}_\Gamma$ is invariant under the action of $\mathcal{O}_{\mathrm{inv}}$, so that the restriction $\mathcal{O}_{\mathrm{inv}}: \mathcal{K}_\Gamma \to \mathcal{K}_\Gamma$ is well defined. Since physical states are required to be invariant under graph automorphisms, physical observables must also be invariant under graph automorphisms. The factor of $1/|\mathrm{Aut }(\Gamma)|$ can be dropped in Eq.~\eqref{eq:O-inv}, and the result is still an invariant observable.

As a direct consequence of Eq.~\eqref{eq:O-inv}, the averages of the operators $\mathcal{O}$ and $\mathcal{O}_{\mathrm{inv}}$ are the same for an invariant state described by a density matrix $\rho^{\mathrm{inv}}_\Gamma$,
\begin{align}
\label{eq:invariant-exp}
\mathrm{Tr}(\rho^{\mathrm{inv}}_\Gamma \, \mathcal{O}_{\mathrm{inv}}) &= \frac{1}{| \mathrm{Aut}(\Gamma)|}  \, \sum_A \, \mathrm{Tr}(\rho^{\mathrm{inv}}_\Gamma \, U_A \, \mathcal{O} \,U_A^{-1}) \nonumber \\
& = \frac{1}{| \mathrm{Aut}(\Gamma)|} \, \sum_A \, \mathrm{Tr}(\rho^{\mathrm{inv}}_\Gamma \, \mathcal{O}) \nonumber \\
& = \mathrm{Tr}(\rho^{\mathrm{inv}}_\Gamma \, \mathcal{O}) \, .
\end{align}

\subsection{Automorphism-invariant states: Bell-network states}
\label{sec:Bell-states}

Bell-network states~\cite{bellnetwork} are a special class of entangled states in loop quantum gravity that maximize correlations of neighboring quantum polyhedra. For large spins, they describe a superposition of vector geometries, a collection of polyhedra glued together so that normals of glued faces are back-to-back, even though in general the faces do not have the same shape. This class of geometries plays an essential role in the study of the asymptotic behavior of topological $SU(2)$ spinfoam vertex amplitudes~\cite{bf1,bf8,bf9,bf10,bf15,bf16}. In this section, we will show that Bell states are autorphism-invariant states on any graph.

Bell-network states are defined by exploring the technique of squeezed vacua in the bosonic representation of loop quantum gravity~\cite{squeezedvacua, bosonic,squeezed-lattice}. In this representation, the Hilbert space $\mathcal{H}_{\Gamma}$ of gauge-invariant states on a graph $\Gamma$ corresponds to a subspace of a bosonic Hilbert space $\mathcal{H}^{\otimes 4L}_{osc}$ of $4L$ harmonic oscillators, where each link $\ell$ is equipped with a pair of oscillators $a^{\alpha \, \dagger}_{s(\ell)}$ at its source node $s(\ell)$ and a pair of oscillators $a^{\alpha \, \dagger}_{t(\ell)}$ at its target node $t(\ell)$,  where $\alpha=1,2$~\cite{bosonic}. To each link there corresponds then a Hilbert space $\mathcal{H}^4_{osc}$ of four oscillators. We denote the projector onto the subspace $\mathcal{H}_{\Gamma}$ by $P_\Gamma:\mathcal{H}_{bos} \to \mathcal{H}_{\Gamma}$.

A Bell state is first defined at a link $\ell$ through
\begin{equation}
\ket{\lambda_\ell, \mathcal{B}} = \left( 1- \left| \lambda_\ell\right|^2 \right) \exp \left( \lambda_\ell \epsilon_{\alpha \beta} \, a^{\alpha \, \dagger}_{s(\ell)} \,\, a^{\beta \, \dagger}_{t(\ell)} \right) \ket{0}_s \ket{0}_t \ ,
\label{eq:link-Bell-state}
\end{equation}
where the squeezing parameter $\lambda_\ell$ is a complex number that encodes the average area and the average extrinsic angle associated with the link~\cite{bellnetwork}. The Bell-network state on a graph $\Gamma$ is the gauge-invariant projection of the tensor product of Bell states at all links,
\begin{equation}
\ket{\Gamma, \{\lambda_\ell\} ,\mathcal{B}} = P_\Gamma \bigotimes_{\ell \in \Gamma} \ket{\lambda_{\ell}, \mathcal{B}} \ .
\end{equation}
The projection can be implemented using the resolution of the identity in the spin-network basis,
\be
P_\Gamma=\sum_{j_\ell, i_n} \ket{\Gamma, \{j_\ell\}, \{i_n\}} \bra{\Gamma, \{j_\ell\}, \{i_n\}} \, ,
\ee
leading to a formula for the Bell-network states as an expansion over spin configurations,
\begin{equation}
\label{BNdefG}
\ket{\Gamma, \{\lambda_\ell\} ,\mathcal{B}} = \sum_{j_\ell} \, \prod_\ell \, q_{j_\ell}(\lambda_\ell) \, \ket{\Gamma, \{j_\ell\},\mathcal{B}}
\end{equation} 
with expansion coefficients
\be
q_{j_\ell}(\lambda_\ell)  = \left( 1- \left| \lambda_\ell\right|^2 \right) \lambda_\ell^{2 j_\ell}  \sqrt{2 j_\ell +1}
\ee
and Bell-network states at fixed spin configuration given by
\begin{equation}
\label{BNdef}
\ket{\Gamma, \{j_\ell\},\mathcal{B}} =\frac{1}{\sqrt{\mathcal{N}}} \, \sum_{i_n} \, \overline{\mathcal{A}_\Gamma \left(j_\ell, i_n \right)} \,\, \ket{\Gamma, \{j_\ell\}, \{i_n\}} \, .
\end{equation}
The quantity $\mathcal{N}$ is a normalization constant, and the amplitude $\mathcal{A}$ is the $SU(2)$-symbol of the graph $\Gamma$,
\be
\label{BNcoeff}
\mathcal{A}_\Gamma \left(j_\ell, i_n \right) = \sum_{\left\lbrace m\right\rbrace} \prod_n [i_n]^{\,\, m_{n1} \cdots m_{n \mu}}{}_{\, m_{n (\mu+1)} \cdots } \, ,
\ee
where the indices of the intertwiner tensors $\left[i_n \right]$ are contracted according to the combinatorics of the graph $\Gamma$.

We wish to prove that Bell-network states $\ket{\Gamma, \{\lambda_\ell\} ,\mathcal{B}}$ with the same squeezing parameter at all links, $\lambda_\ell=\lambda$, are invariant under the action of automorphisms on any graph $\Gamma$. We denote these states by $\ket{\Gamma,\lambda,\mathcal{B}}$. We start by proving a basic property of the Bell states $\ket{\lambda_\ell, \mathcal{B}}$ defined at the links of the graph by Eq.~\eqref{eq:link-Bell-state}. It is convenient to use the holonomy representation for this purpose. At each link $\ell$, states in the bosonic representation $\mathcal{H}$ are mapped to the holonomy representation through the identification
\be
(-1)^{j_\ell-m_\ell} \ket{j_\ell,m_\ell}_{t(\ell)} \ket{j_\ell,-n_\ell}_{s(\ell)} \mapsto \sqrt{2j_\ell+1} [D^{j_\ell}(h_\ell)]^{m_\ell}{}_{n_\ell} \, ,
\ee
where the spin states in the magnetic number basis are defined at any endpoint of a link in terms of the corresponding pair of oscillators by
\be
\ket{j,m} = \frac{\left( a^{0 \, \dagger} \right)^{j+m} \left( a^{1 \, \dagger} \right)^{j-m}}{\sqrt{(j+m)!(j-m)!}}\ket{0} \, .
\ee
The Bell state at a link with squeezing parameter $\lambda_\ell=\lambda$ assumes the following form in the holonomy representation:
\be
\label{Bellstateholo}
\psi(h_\ell) = \scalar{h_\ell}{\lambda,\mathcal{B}} = \left( 1- \left| \lambda\right|^2 \right) \sum_{j_\ell} \lambda^{2 j_\ell} \sqrt{2 j_\ell +1}  \, \Tr[ D^{j_\ell}(h_\ell)] \, .
\ee
The $SU(2)$ character $\Tr[ D^{j_\ell}(h_\ell)]$ is invariant under inversion of its argument. As a result, the state has the symmetry $\psi(h_\ell)=\psi(h^{-1}_\ell)$. Therefore, the state is invariant under orientation reversal.

We can now show that the tensor product 
\be
\Psi(h_{\ell_1}, \dots,h_{\ell_L}) = \psi(h_{\ell_1}) \cdots \psi(h_{\ell_L})
\ee
of Bell states with the same squeezing parameter $\lambda_\ell = \lambda$ at all links of a graph is invariant under the action of an automorphism:
\begin{align*}
\left( U_A \Psi \right)(h_{\ell_1},\dots,h_{h_{\ell_L}}) &= \Psi(h_{A(\ell_1)},\dots,h_{A(\ell_L)}) \\
	&= \prod_{\ell=1}^L \psi(h_{A(\ell_i)}) \\
	&= \prod_{\ell=1}^L \psi(h_{\ell_i}) = \Psi(h_{\ell_1},\dots,h_{h_{\ell_L}}) \,,
\end{align*}
where we used the fact that $\psi(h_{A(\ell_i)})=\psi(h_{\ell_j})$ for some $j$, regardless of whether $A(\ell_i)$ equals $\ell_j$ or the link $\ell^{-1}_j$ with the reverse orientation, due to the symmetry $\psi(h_\ell)=\psi(h^{-1}_\ell)$, and that the automorphism produces a permutation of the unoriented links. We conclude that
\be
U_A \bigotimes_{\ell \in \Gamma} \ket{\lambda, \mathcal{B}}=\bigotimes_{\ell \in \Gamma} \ket{\lambda, \mathcal{B}} \, .
\label{UA-bell-no-proj}
\ee

In addition, the orthogonal projector $P_\Gamma$ onto the subspace of gauge-invariant states is automorphism-invariant. Indeed, for any gauge-invariant $\Psi$, its image $U_A \Psi$ under an automorphism $A$ is also gauge-invariant:
\begin{align}
(U_A \Psi)\left(U_{s(\ell)} h_{\ell} U_{t(\ell)}\right) &= \Psi\left( U_{s(A(\ell))} h_{A(\ell)} U_{t(A(\ell))} \right) \nonumber \\
	&= \Psi( h_{A(\ell)} ) \nonumber \\
	&= (U_A \Psi)(h_\ell) \,.
\end{align}
As the automorphism is invertible and its inverse is an autormorphism, any gauge-invariant state is in the image of $U_A$. Therefore, $\mathcal{H}_\Gamma$ is mapped onto itself by $U_A$. Since $U_A$ is unitary, the orthogonal complement of $\mathcal{H}_\Gamma$ is also mapped onto itself. It follows that
\be
U_A P_\Gamma U_A^{-1} = P_\Gamma \, .
\label{UA-preserve-gauge-inv}
\ee

The invariance of the Bell-network states under automorphisms now follows directly from Eqs.~\eqref{UA-bell-no-proj} and \eqref{UA-preserve-gauge-inv}:
\begin{align*}
U_A \ket{\Gamma, \lambda ,\mathcal{B}} &= U_A P_\Gamma \bigotimes_{\ell \in \Gamma} \ket{\lambda, \mathcal{B}} \\
	&= U_A P_\Gamma U_A^{-1} U_A \bigotimes_{\ell \in \Gamma} \ket{\lambda, \mathcal{B}} \\
	&= P_\Gamma \bigotimes_{\ell \in \Gamma} \ket{\lambda, \mathcal{B}} \\
	&= \ket{\Gamma, \lambda ,\mathcal{B}} \, ,
\end{align*}
completing the proof.

\subsection{Homogeneity and isotropy}

As discussed in Sec.~\ref{sec:decompositions}, $1$-CH and $2$-CH graphs are associated with regular decompositions of homogeneous and isotropic spaces. Their symmetries mirror at a combinatorial level the symmetries of homogeneous and isotropic geometries in the continuum. In this section, we discuss states and observables invariant under graph automorphisms on $k$-CH graphs and describe their properties of homogeneity and isotropy.

\subsubsection{$1$-CH graphs}
\label{sec:1-ch}

Let $\Gamma_H$ be a $1$-CH graph. A state $\ket{\Psi_{H}} \in  \mathcal{K}_{\Gamma_H} \subset \mathcal{H}_{\Gamma_H}$ can be written in the spin-network basis as
\be
\ket{\Psi_{H}} = \sum \limits_{j_\ell, i_n}  c_{j_{\ell}, i_n} \, \ket{\Gamma_H, \{j_{\ell}\}, \{i_n\}} \, ,
\ee
and satisfies $U_A \ket{\Psi} = \ket{\Psi}$, $\forall A \in \mathrm{Aut}(\Gamma_H)$, where the sums over spins run over $j_\ell=1/2,1,\dots$, and the sums over intertwiner indices run over $i_n=1,\dots, \dim \mathcal{H}_n$. Let its density matrix be represented by
\[
\rho_H = \ket{\Psi_H} \bra{\Psi_H} \, .
\]

In this section, we will construct observables on $\mathcal{K}_{\Gamma_H}$ that represent measurements performed on a single node and respect the symmetry under automorphisms of the theory. As all nodes are equivalent in a $1$-CH graph, this cannot be done by taking an operator that acts nontrivially on a single node, which would break automorphism-invariance. In fact, we are led to introduce one-node observables that are similar in nature to the one-body operators used in many-body quantum mechanics \cite{martin-rothen}. We also introduce a reduced density matrix that describes the statistics of one-node observables, which will later be used for the definition of the entanglement entropy of a single node on the graph.

\paragraph*{Local observables.}

If $r \in N(\Gamma_H)$ is a node of the graph $\Gamma_H$, we say that an operator $\mathcal{O}_r:\mathcal{H}_{\Gamma_H} \to \mathcal{H}_{\Gamma_H}$ is a local one-node operator at the node $r$ if its action on the spin-network basis has the form:
\begin{align}
\mathcal{O}_r \ket{\Gamma_H, \{j_{\ell}\}, \{i_n\}} 
	= \sum_i \mathcal{O}_{i_r,i} \ket{\Gamma_H, \{j_{\ell}\}, \{i_1, \dots, i_{r-1}, i, i_{r+1}, \dots, i_N\}} \, .
\label{eq:local-operator}
\end{align}
Such an operator preserves the spins and, for any spin configuration, has a nontrivial action only in the intertwiner space $\mathcal{H}_r$ associated with the node $r$. Its restriction to a subspace of fixed spins has the form:
\be
\mathcal{O}_r{}|_{\mathcal{H}_{\Gamma_H, \{j_\ell\}}} = \mathds{1} \otimes \cdots  \otimes \mathcal{O}_{\{j_{ra}\}} \otimes \cdots \otimes \mathds{1} \, ,
\label{eq:local-operator-2}
\ee
for some operator $\mathcal{O}_{\{j_{ra}\}}:\mathcal{H}_r \to \mathcal{H}_r$, where $j_{r1},\dots,j_{rV}$ are the spins at the node $r$. A local one-node operator is completely characterized by the list of operators $\mathcal{O}_{\{j_{ra}\}}$ for all possible spin configurations $\{j_{ra}\}$ at the node.

Under an automorphism $A$ for which $A(r)=s$, the operator $\mathcal{O}_r$ transforms as:
\begin{align*}
U_A \mathcal{O}_r U_A^{-1} \ket{\Gamma_H, \{j_{\ell}\}, \{i_n\}} & =  s_{A^{-1},j_\ell} U_A \mathcal{O}_r \ket{\Gamma_H, \{j'_\ell\}, \{i'_n\}} \\
	&= s_{A^{-1},j_\ell} U_A \sum_{k} \mathcal{O}'_{i'_r,k} \ket{\Gamma_H, \{j'_\ell \}, \{\dots,i'_{r-1},k,i'_{r+1},\dots\}} \\
	&= \sum_k \mathcal{O}_{i_s,k} \ket{\Gamma_H, \{j_\ell\}, \{ \dots,i_{s-1},k,i_{s+1},\dots\}} \\
	&= \mathcal{O}_s \ket{\Gamma_H, \{j_{\ell}\}, \{i_n\}} \, .
\end{align*}
In the second line, $i'_n=i_{A(n)}$. The matrix $\mathcal{O}'$ describes the operator $\mathcal{O}_r$ in the basis $\ket{\Gamma_H, \{j'_\ell\}, \{i'_n\}}$. In the passage from the second to the third line, it was used that $i'_r=i_s$, and one can recognize the action of a local operator at the node $s$, which we denote by $\mathcal{O}_s$, with the same matrix elements as $\mathcal{O}'$. We have found that
\be
U_A \mathcal{O}_r U_A^{-1} = \mathcal{O}_s \, ,
\label{eq:transformation-local-observable}
\ee
i.e., the local operator at the node $r$ is transformed into a local operator at the node $s=A(r)$.

It follows from Eq.~\eqref{eq:transformation-local-observable} and the definition \eqref{eq:O-inv} that the invariant observables associated with a local operator $\mathcal{O}_r$ and its image $\mathcal{O}_s$ under an automorphism are the same:
\[
\mathcal{O}_{r,\text{inv}} = \mathcal{O}_{s,\text{inv}} \equiv \mathcal{O}^1_{\text{inv}}\, .
\]
Now, from Eq.~\eqref{eq:invariant-exp}, we find that the averages of local operators $\mathcal{O}_r,\mathcal{O}_s$ agree, for any invariant state $\rho^{\text{inv}}_{\Gamma_H}$:
\be
\Tr (\rho^{\text{inv}}_{\Gamma_H} \, \mathcal{O}_r) = \Tr (\rho^{\text{inv}}_{\Gamma_H} \, \mathcal{O}^1_{\text{inv}}) = \Tr (\rho^{\text{inv}}_{\Gamma_H} \, \mathcal{O}_s)   \, .
\label{eq:local-to-inv}
\ee
We call the invariant operator $\mathcal{O}_{r,\text{inv}}$ associated with a local one-node operator $\mathcal{O}_r$ an invariant one-node operator.

The operator $\mathcal{O}^1_{\text{inv}}:\mathcal{K}_{\Gamma_H}\to\mathcal{K}_{\Gamma_H}$ is the adequate tool to describe the measurement of a local property of the geometry at a single node in the space of automorphism-invariant states. The node at which the measurement is performed is not specified by the operator, due to the group-averaging over the actions of the automorphisms, which makes the operator nonlocal with respect to the graph. As all nodes are equivalent in a $1$-CH graph, any operator that distinguishes between nodes cannot in fact be an observable on the space $\mathcal{K}_{\Gamma_H}$ of automorphism-invariant states. Such a nonlocality becomes more evident in an explicit representation of the invariant operator as an average of local one-node operators over the graph, which we discuss now for the case of $1$-CH graphs.

We have seen that a local operator $\mathcal{O}_m$ at a node $m$ can be associated with a given local operator $\mathcal{O}_n$ at a node $n$ through Eq.~\eqref{eq:transformation-local-observable}. This association is not unique, however, if there is more than one automorphism relating the nodes. Suppose, for instance, that a particular component of the Penrose metric, say $g_{12}$, is measured at $n$. It might happen that this operator transforms under an automorphism into an operator $g_{12}$ at $m$, while transforming into $g_{23}$ under another automorphism. However, if the operator $\mathcal{O}_n$ is invariant under automorphisms that preserve $n$, its image at $m$ is always the same, for any automorphism relating the two nodes, as we will show now. Examples of such observables are the total area and the volume at a node.

Consider a one-node local observable $\mathcal{O}_n$ that is invariant under automorphisms that preserve $n$, and let $A,A'$ be automorphisms for which $A(n)=A'(n)=m$. Denote its transformation under the automorphisms by
\begin{align}
U_A \mathcal{O}_n U_A^{-1} &= \mathcal{O}_m \, ,\nonumber \\
U_{A'} \mathcal{O}_n U_{A'}^{-1} &= \mathcal{O}'_m \, .
\label{eq:aut-invariant-at-n}
\end{align}
The composition $(A')^{-1} \circ A$ is an automorphism that preserves the node $n$. Therefore,
\begin{align*}
\mathcal{O}_n &= (U_{A'}^{-1} U_A) \mathcal{O}_n (U_{A'}^{-1} U_A)^{-1} \\
	&= U_{A'}^{-1} \mathcal{O}_m U_{A'}
\end{align*}
This implies that
\[
U_{A'} \mathcal{O}_n U_{A'}^{-1} = \mathcal{O}_m \, ,
\]
which, together with Eq.~\eqref{eq:aut-invariant-at-n}, leads to $\mathcal{O}_m=\mathcal{O}'_m$.

Let the set of automorphisms that preserve the node $n$ be denoted by $[\mathrm{Aut}\; (\Gamma_H)]_n$ and the set of automorphisms that map the node $n$ to the node $m$ by $[\mathrm{Aut }\; (\Gamma_H)]_{n\to m}$. From Eqs.~\eqref{eq:O-inv} and \eqref{eq:transformation-local-observable}, we have then:
\begin{align*}
\mathcal{O}^1_{\mathrm{inv}} &= \frac{1}{|\mathrm{Aut }(\Gamma_H)|} \,\, \sum_A U_A \, \mathcal{O}_n \, U_A^{-1}  \\
	&= \frac{1}{|\mathrm{Aut }(\Gamma_H)|} \,\, \sum_A \, \mathcal{O}_{A(n)} \\
	&= \frac{1}{|\mathrm{Aut }(\Gamma_H)|} \,\, \sum_m \, \left| [\mathrm{Aut }\; (\Gamma_H)]_{n\to m} \right| \mathcal{O}_m \, .
\end{align*}
We claim that $\left| [\mathrm{Aut }\; (\Gamma_H)]_{n\to m} \right| = \left| [\mathrm{Aut}\; (\Gamma_H)]_n \right|$. Let us enumerate the automorphisms in $[\mathrm{Aut }\; (\Gamma_H)]_{n\to m}$ as $f_1, \dots, f_p$ and the automorphisms in $[\mathrm{Aut}\; (\Gamma_H)]_n$ as $h_1,\dots, h_q$. We first note that $f_i=f_1 \circ (f_1^{-1} \circ f_i)$, where $f_1^{-1} \circ f_i \in [\mathrm{Aut}\; (\Gamma_H)]_n$. Therefore, any element of $[\mathrm{Aut }\; (\Gamma_H)]_{m\to n}$ can be written as the composition of an element of $[\mathrm{Aut}\; (\Gamma_H)]_n$ and a fixed automorphism $f_1$. Moreover, if $f_i \neq f_j$, then $f_1^{-1} \circ f_i  \neq f_1^{-1} \circ f_j$, since $f_1^{-1}$ is invertible. Hence, $\left| [\mathrm{Aut }\; (\Gamma_H)]_{n\to m} \right| \leq \left| [\mathrm{Aut}\; (\Gamma_H)]_n \right|$. Moreover, for any $h_j$, the composition $f_1 \circ h_j$ is in $[\mathrm{Aut }\; (\Gamma_H)]_{n\to m}$. If $h_k \neq h_l$, then $f_1 \circ h_k \neq f_1 \circ h_l$, since $f_1$ is invertible. Therefore, $\left| [\mathrm{Aut }\; (\Gamma_H)]_{n\to m} \right| \geq \left| [\mathrm{Aut}\; (\Gamma_H)]_n \right|$, and the result follows. 

A direct consequence of this result is that $\left| [\mathrm{Aut }\; (\Gamma_H)]_{n\to m} \right|$ is independent of $m$. Moreover, since $\cup_m [\mathrm{Aut }\; (\Gamma_H)]_{n\to m} = \mathrm{Aut }\; (\Gamma_H)$, it follows that $\left| [\mathrm{Aut }\; (\Gamma_H)]_{n\to m} \right| = |\mathrm{Aut }\; (\Gamma_H)|/N$. Therefore,
\be
\mathcal{O}^1_{\mathrm{inv}} = \frac{1}{N} \sum_m \mathcal{O}_m \, .
\label{eq:one-node-obs}
\ee
The invariant observable is just the average of the local operators over all nodes of the graph, as could be expected. This is true for $1$-CH graphs, but not for general graphs, in which there can be no automorphism relating a given pair of nodes.

The formula \eqref{eq:one-node-obs} for the invariant observable associated with a local one-node is reminiscent of the definition of one-body operators in many-body quantum mechanics \cite{martin-rothen}. In our approach, such a representation of invariant one-node observables appears naturally for the description of local measurements performed at a single node on $1$-CH graphs, respecting the invariance under automorphisms, and for local one-node operators that are invariant under automorphisms that preserve the node. Removing the factor of $1/N$ in Eq.~\eqref{eq:one-node-obs}, we obtain another invariant observable
\be
\mathcal{O}^1_{\mathrm{inv},T} = \sum_m \mathcal{O}_m
\label{eq:one-node-obs-2}
\ee
that describes the total sum of the local operators over the nodes, instead of their average.

The result can be extended to local one-node observables that are not invariant under automorphisms that preserve the node. In the case of a local operator $\mathcal{O}_n$ that can be mapped into distinct local operators $\mathcal{O}_m, \mathcal{O}'_m, \dots$ at $m$ by distinct automorphisms, one can still define a one-node observable through \eqref{eq:one-node-obs} by choosing a particular representation $\mathcal{O}_m$ at each node $m$. The matrix elements of the resulting operator are independent of this choice for states on $\mathcal{K}_{\Gamma_H}$, since the matrix elements of $\mathcal{O}_m=U_A \mathcal{O}_n U_A^{-1}$ and $\mathcal{O}'_m=U_{A`} \mathcal{O}_n U_{A'}^{-1}$ are the same for such states:
\begin{align*}
P_A \mathcal{O}_m P_A &= P_A U_A \mathcal{O}_n U_A^{-1} P_A  \\
&= P_A \mathcal{O}_n P_A  \, ,
\end{align*}
and, similarly, $P_A \mathcal{O}'_m P_A = P_A \mathcal{O}_n P_A$, showing that $P_A \mathcal{O}_m P_A=P_A \mathcal{O}'_m P_A$. As a result, we can express the invariant observable associated with $\mathcal{O}_n$ as the restriction $P_A \mathcal{O}^1_{\mathrm{inv}} P_A$ with $\mathcal{O}^1_{\mathrm{inv}}$ given by Eq.~\eqref{eq:one-node-obs}, for an arbitrary choice of representation $\mathcal{O}_m$ for the image of the local operator at each node.

A simple example of a one-node observable is the total area $J_H$ at a node,
\be
J_H =  \frac{1}{N} \, \sum_n \mathds{1} \otimes \cdots  \otimes J_n \otimes \cdots \otimes \mathds{1}\,,
\ee
where the action of $J_n$ at the node $n$ is given by
\begin{align}
J_n = \sum \limits_{a=1}^V \sqrt{g_{aa}(n)}  \,.
\end{align}
Another example is the volume $V_H$ at a node, 
\be
V_H =  \frac{1}{N} \, \sum_n \mathds{1} \otimes \cdots  \otimes V_n \otimes \cdots \otimes \mathds{1}\,,
\label{eq:volume-operator-inv}
\ee
where $V_n$ is the volume operator at the node $n$ on $\mathcal{H}_{\Gamma_H}$. For a $4$-valent node, for instance, the volume operator can be expressed in terms of the Penrose metric as
\be
V_n =  \frac{\sqrt{2}}{3} \sqrt{ \big| \textstyle{\frac{i}{\mathfrak{a}_{0}}} \, [g_{ab}(n), g_{ac}(n)] \, \big|} \,,
\label{eq:volume-operator-node}
\ee
where $a,b,c$ label any three distinct links at the node. Volume operators can also be defined for nodes of arbitrary valency \cite{rsvolume,alvolume,bianchi-polyhedra}. Distinct versions of the volume operator have been constructed, due to regularization ambiguities, but invariant observables can be associated with any of them, as far as the volume operator is a local one-node observable, which is true for both the Rovelli-Smolin \cite{rsvolume} and the Ashtekar-Lewandowski operators \cite{alvolume}, as well as for that proposed by Bianchi in \cite{bianchi-polyhedra}. The total volume is also an invariant observable:
\be
V_{H,T} =  \sum_n \mathds{1} \otimes \cdots  \otimes V_n \otimes \cdots \otimes \mathds{1}\,,
\ee

Invariant observables describing measurements performed on a pair of nodes can be similarly constructed,
\begin{align*}
\mathcal{O}^2_{mn,\mathrm{inv}} &= \frac{1}{|\mathrm{Aut }(\Gamma_H)|} \,\, \sum_A U_A \, \mathcal{O}_m \mathcal{O}_n \, U_A^{-1}  \nonumber \\
	&= \frac{1}{|\mathrm{Aut }(\Gamma_H)|} \,\, \sum_A \, \mathcal{O}_{A(m)} \mathcal{O}_{A(n)}  \, .
\end{align*}
From Eq.~\eqref{eq:invariant-exp}, their averages on automorphism-invariant states are given by
\[
\Tr (\rho^{\mathrm{inv}}_{\Gamma_H} \mathcal{O}^2_{mn,\mathrm{inv}}) = \Tr (\rho^{\mathrm{inv}}_{\Gamma_H} \mathcal{O}_m \mathcal{O}_n) \, .
\]
When the local operators $\mathcal{O}_m,\mathcal{O}_n$ describe the same physical quantity at the distinct nodes, the observable $\mathcal{O}^2_{\mathrm{inv}}$ describes its correlation function through
\begin{align*}
\langle \mathcal{O} \mathcal{O}\rangle_{mn} &= \Tr (\rho^{\mathrm{inv}}_{\Gamma_H} \mathcal{O}^2_{mn,\mathrm{inv}}) - \Tr (\rho^{\mathrm{inv}}_{\Gamma_H} \mathcal{O}^1_{m,\mathrm{inv}}) \Tr (\rho^{\mathrm{inv}}_{\Gamma_H} \mathcal{O}^1_{n,\mathrm{inv}}) \\
	&= \Tr (\rho^{\mathrm{inv}}_{\Gamma_H} \mathcal{O}^2_{mn,\mathrm{inv}}) - \Tr (\rho^{\mathrm{inv}}_{\Gamma_H} \mathcal{O}^1_{\mathrm{inv}})^2 \, , 
\end{align*}
where we used the fact that, for any pair of nodes $m,n$ in a $1$-CH graph, the invariant one-node observables satisfy $\mathcal{O}^1_{m,\mathrm{inv}}=\mathcal{O}^1_{n,\mathrm{inv}} = \mathcal{O}^1_{\mathrm{inv}}$.

It is not possible to map any two pairs of nodes into each other by some automorphism in a $1$-CH graph. As a result, the observable $\mathcal{O}^2_{mn,\mathrm{inv}}$ does not admit a simplification analogous to the formula \eqref{eq:one-node-obs}. Nevertheless, in analogy with two-body operators in quantum mechanics, we can define invariant two-node physical observables as
\be
\mathcal{O}^2_{\mathrm{inv}} = \frac{2}{N(N-1)} \,\, \sum \limits_{m \neq n} \,\mathcal{O}_m \mathcal{O}_n \, ,
\label{eq:two-node-observable}
\ee
for any local operator $\mathcal{O}_n$ that is invariant under automorphims that preserve the node $n$. The extension to multiple-node invariant observables $\mathcal{O}^r_{\mathrm{inv}}$ for $r >2$ is immediate. The two-node observable \eqref{eq:two-node-observable} is the average of the invariant observables $\mathcal{O}^2_{mn,\mathrm{inv}}$,
\[
\mathcal{O}^2_{\mathrm{inv}} = \frac{2}{N(N-1)} \,\, \sum \limits_{m \neq n} \, \mathcal{O}^2_{mn,\mathrm{inv}} \, .
\]

\paragraph*{Invariant one-node reduced density matrix.}

We can introduce a reduced density matrix $\rho^1$ that describes the statistics of invariant one-node observables on a $1$-CH graph as follows. Let $r$ be an arbitrary node, and consider a local one-node operator of the form \eqref{eq:local-operator}, with $\mathcal{O}^1_{\mathrm{inv}}$ its associated invariant observable. The orientations of all links at the node $r$ can be chosen to point outwards from the node. Each of the orthogonal subspaces $\mathcal{H}_{\Gamma_H, \{j_\ell\}}$ of fixed spins is invariant under the action of local operators. As a result, the average of any one-node observable has the form
\[
\mean{\mathcal{O}^1_{\mathrm{inv}}} = \mean{\mathcal{O}_r} = \sum_{\{j_{ra} \}} p_{\{j_{ra} \}} \Tr \left(\mathcal{O}_{\{j_{ra}\}}  \rho_{\{j_{ra}\}} \right) \, ,
\]
with
\be
p_{\{j_{ra} \}} = \sum_{\{j_\ell  \neq j_{ra}\}} \sum_{\{i_n\}} |c_{j_{\ell}, i_n}|^2 \, ,
\label{eq:node-spins-prob}
\ee
where the first sum runs over all spins, except those at the node $r$, which are fixed with values $j_{ra}=j_a$, the second sum runs over all intertwiner indices, and
\begin{align}
\rho_{\{j_{ra}\}} =  \frac{1}{p_{\{j_{ra} \}}} \sum_{\{j_\ell  \neq j_{ra}\}} \sum_{\{i_n \neq i_r\}} \sum_{i,\bar{i}} c_{j_{\ell}, \{i_1,\dots,i,\dots,i_N\}} \, c^*_{j_{\ell}, \{i_1,\dots,\bar{i},\dots,i_N\}} \ket{i} \bra{\bar{i}} \, ,
\label{eq:node-reduced-rho}
\end{align}
where the sums over $i,\bar{i}$ run over all orthogonal intertwiners of $\mathrm{Inv}_{\mathrm{SU(2)}} \left[  \bigotimes_a \, \mathcal{V}_{j_a}  \right]$.

The reduced density matrix $\rho_{\{j_{ra}\}}$ is an operator on the intertwiner space $\mathcal{H}_r$ with spins $j_{ra}=j_a$ associated with the node $r$. Now, for an automorphism-invariant state, the same density matrix is obtained at any node $s$, if the links at $s$ are ordered so that $A(ra)=(sa)$, for some automorphism that takes $r$ to $s$, and if the same spin configuration is considered, $j_{sa}=j_a$. To prove it, we first note that, for any invariant state
\be
\ket{\Psi} = \sum \limits_{j_\ell, i_n}  c_{j_{\ell}, i_n} \, \ket{\Gamma_H, \{j_{\ell}\}, \{i_n\}} \, ,
\label{eq:automorphism-spin-network-basis-1}
\ee
it follows from Eq.~\eqref{eq:autoaction} that
\be
\ket{\Psi} = U_A \ket{\Psi_H} = \sum \limits_{j'_\ell, i'_n} s_{A,j_\ell} c'_{j'_{\ell}, i'_n} \, \ket{\Gamma_H, \{j'_{\ell}\}, \{i'_n\}} \, ,
\label{eq:automorphism-spin-network-basis}
\ee
with $c'_{j'_{\ell}, i'_n}=c_{j_{\ell}, i_n}$, for $j'_\ell = j_{A^{-1} \ell} \, , i'_n = i_{A^{-1}n}$. Fixing $j'_{sa}=j_a$, we have
\begin{align*}
p_{\{j_{sa} \}} &= \sum_{\{j'_\ell  \neq j'_{sa}\}} \sum_{\{i'_n\}} |c'_{j'_{\ell}, i'_n}|^2  \, , \quad \text{with } j'_{sa}=j_a \\
&= \sum_{\{j_\ell  \neq j_{ra}\}} \sum_{\{i_n\}} |c_{j_{\ell}, i_n}|^2 \, , \quad \text{with } j_{ra}=j_a\\
&= p_{\{j_{ra} \}} \, .
\end{align*}
A similar argument shows that $\rho_{\{j_{sa}\}}$ and $\rho_{\{j_{ra}\}}$ have the same matrix elements:
\begin{align}
\bra{\kappa} \rho_{\{j_{sa}\}} \ket{\tilde{\kappa}} &=  \frac{1}{p_{\{j_{sa} \}}} \sum_{\{j'_\ell  \neq j'_{sa}\}} \sum_{\{i'_n \neq i'_s\}} c'_{j'_{\ell}, \{i'_1,\dots,\kappa,\dots,i'_N\}}  c'^*_{j'_{\ell}, \{i'_1,\dots,\tilde{\kappa},\dots,i'_N\}} \nonumber \nonumber \\
&=  \frac{1}{p_{\{j_{ra} \}}} \sum_{\{j_\ell  \neq j_{ra}\}} \sum_{\{i_n \neq i_r\}} c_{j_{\ell}, \{i_1,\dots,\kappa,\dots,i_N\}}  c^*_{j_{\ell}, \{i_1,\dots,\tilde{\kappa},\dots,i_N\}}  \nonumber \nonumber \\
&= \bra{\kappa} \rho_{\{j_{ra}\}} \ket{\tilde{\kappa}} \, ,
\label{eq:homog-same-rho1}
\end{align}
where the indices $\kappa,\tilde{\kappa}$ refer to the intertwiner basis $i'_s$ in the first line, and to the intertwiner basis $i_r$ in the second and third lines.

The property expressed in Eq.~\eqref{eq:homog-same-rho1} is perhaps the most direct way to characterize the homogeneity of states in $\mathcal{K}_{\Gamma_H}$: the reduced density matrix $\rho_{\{j_a\}}$ that describes the statistics of observations performed on a single node is the same for all nodes.

From Eq.~\eqref{eq:local-to-inv}, the average of an invariant observable $\mathcal{O}^1_{\mathrm{inv}}$ on a $1$-CH graph can be computed using its associated local one-node operator $\mathcal{O}_n$ at a node $n$, for any node $n$. Hence, all invariant observables can be studied at a common reference node, which we take to be $r$. In a $1$-CH graph, all nodes have the same valency $V$. Let us introduce an abstract intertwiner space
\be
\mathcal{H}^1 = \bigoplus_{\{j_a\}} \mathcal{H}_{\{j_a\}} \, , \quad \mathcal{H}_{\{j_a\}} = \mathrm{Inv}_{\mathrm{SU(2)}} \left[  \bigotimes_{a=1}^V \, \mathcal{V}_{j_a}  \right] \, ,
\label{eq:abstract-intertwiner-space}
\ee
isomorphic to the intertwiner stace at the node $r$, $\mathcal{H}^1 \simeq \mathcal{H}_r$, in order to describe the statistics of invariant one-node observables associated with local operators. A local invariant observable $\mathcal{O}^1_{\text{inv}}$ on $\mathcal{K}_{\Gamma_H}$ is naturally associated with an operator
\be
\mathcal{O}^1 = \bigoplus_{\{j_a\}} \mathcal{O}_{\{j_a\}}
\label{eq:abstract-one-node-obs}
\ee
on $\mathcal{H}^1$, with 
\be
\mathcal{O}_{\{j_a\}}=\mathcal{O}_{\{j_{ra}\}} \, ,
\label{eq:abstract-one-node-obs-2}
\ee
and an automorphism-invariant state is represented by a density matrix $\rho^1: \mathcal{H}^1 \to \mathcal{H}^1$ defined as
\be
\rho^1 = \sum_{\{j_a\}} p_{\{j_a\}} \rho^1_{\{j_a\}} \, ,
\label{eq:abstract-one-node-state}
\ee
which we call the invariant one-node density matrix, where $\rho^1_{\{j_a\}}=\rho_{\{j_{ra}\}}$ and $p_{\{j_a\}}=p_{\{j_{ra}\}}$. Then:
\be
\mean{\mathcal{O}^1_{\mathrm{inv}}} = \Tr_{\mathcal{H}^1} (\rho^1 \mathcal{O}^1) \, .
\label{eq:abstract-one-node-av}
\ee
The average of an invariant one-node observable can thus be calculated as an average on an abstract intertwiner space with orthogonal subspaces labelled by boundary spins $j_a$.

The matrix elements of the density matrix $\rho^1$ and of the observable $\mathcal{O}^1$ depend on the choice of an intertwiner basis at the reference node $r$ used to establish the isomorphism $\mathcal{H}^1 \simeq \mathcal{H}_r$, as well as on the choice of the reference node. We can impose a consistency condition on the intertwiner bases, however, which ensure that $\rho^1$ and $\mathcal{O}^1$ become independent of the choice of reference node. Let $s$ be any node, and $A$ an automorphism such that $A(r)=s$. For a given intertwiner basis $\ket{i_r}$ at the node $r$, there corresponds an intertwiner basis $\ket{i'_s}=U_A\ket{i_r}$ at the node $s$. From \eqref{eq:homog-same-rho1}, the reduced density matrix has the same form at both nodes in these bases. As a result, in such bases, the same invariant one-node density matrix $\rho^1$ is obtained whether $r$ or $s$ is taken as a reference node. Consider now a local operator $\mathcal{O}_r$ at the node $r$ with associated invariant observable $\mathcal{O}^1_{\mathrm{inv}}$. Then $\mathcal{O}_s=U_A \mathcal{O}_r U_A^{-1}$ is a local operator at the node $s$ associated with the same invariant observable. Its matrix elements in the basis $\ket{i'_s}$ are equal to those of $\mathcal{O}_r$ in the basis $\ket{i_r}$,
\[
\bra{i'_s} \mathcal{O}_s \ket{\tilde{i}'_s} = \bra{i'_s} U_A \mathcal{O}_r U_A^{-1} \ket{\tilde{i}'_s} = \bra{i_r} \mathcal{O}_r \ket{\tilde{i}_r} \, .
\]
Hence, the same operator $\mathcal{O}^1$ is associated with $\mathcal{O}^1_{\mathrm{inv}}$ regardless of the choice of the reference node. We require that a consistent choice of intertwiner bases is made for the representation of invariant states and observables in the abstract intertwiner space $\mathcal{H}^1$, so that the construction is independent of the choice of reference node.

The invariant one-node density matrix $\rho^1$ describes the statistics of measurements of the geometry performed on a single node, which remains unspecified. It decomposes into a direct sum of density matrices $\rho^1_{\{j_a\}}$ labelled by the possible boundary spins, and each component $\rho^1_{\{j_a\}}$ describes the statistics of measurements performed within the region delimited by the specified boundary spins. One can ask, for instance, what is the average volume of a node, given that spins $j_a$ were observed at its links, without specifying the node at which the observations are to be made. On the other hand, one cannot ask about the average volume at a specific node, as no invariant operator describing such a measurement exists on $\mathcal{K}_{\Gamma_H}$. Later, we will also explore the density matrix $\rho^1$ as a tool for the description of measurements performed on a single node for states involving a superposition of graphs.

For two-node observables on a $1$-CH graph, there is no direct analogue of the one-node density matrix defined on an abstract space that is independent of the nodes. It is straightforward to introduce a density matrix $\rho^2_{mn}$ analogous to $\rho^1$ following similar steps, but in general it will depend on the choice of the nodes $m,n$. If two pairs $(r,s)$ and $(m,n)$ can be related by an automorphism, then it must be $\rho^2_{mn}=\rho^2_{rs}$, since the invariant observables must then satisfy $\mathcal{O}^2_{mn,\mathrm{inv}}=\mathcal{O}^2_{rs,\mathrm{inv}}$. When this is not the case, the density matrices can differ, and there will be a two-node density matrix for each class of pairs of nodes that cannot be related by any automorphism.

\subsubsection{$2$-CH graphs}

Let $\Gamma_C$ be a $2$-CH graph, and $\mathcal{K}_{\Gamma_C} \subset \mathcal{H}_{\Gamma_C}$ be the space of automorphism-invariant states on this graph. Any $2$-CH graph is also $1$-CH, hence the states $\ket{\Psi_C} \in \mathcal{K}_{\Gamma_C}$ are homogeneous and all nodes are equivalent. In addition, in a $2$-CH graph, any two oriented links $\ell$ and $\ell'$ are also equivalent. This allows the notion of local one-link observables to be introduced, in analogy with local one-node observables. As any two links at a node can be related by an automorphism that preserves the node, the invariant states display a discrete version of the property of isotropy: distinct links at a node describe the directions from the node in the graph, and local measurements cannot distinguish among the links.

In the holonomy representation, a generic state $\ket{\Psi_C} \in \mathcal{K}_{\Gamma_C}$ is represented by a wavefunction $\Psi_C(h_\ell)$ that is invariant under gauge transformations and automorphisms. The space of automorphism invariant states is a subspace $\mathcal{K}_{\Gamma_C} \subset \mathcal{H}_{\Gamma_C} \subset \bigotimes_\ell \mathcal{H}_\ell$, where $\mathcal{H}_\ell=L^2[SU(2)]$ is a local Hilbert space associated with the link $\ell$. We say that an operator $\mathcal{O}_\ell:\mathcal{H}_{\Gamma_C} \to \mathcal{H}_{\Gamma_C}$ is a local link operator if it acts nontrivially only on a single link:
\[
\mathcal{O}_\ell = \mathds{1} \otimes \cdots  \otimes \mathcal{O} \otimes \cdots \otimes \mathds{1}\,,
\]
where $\mathcal{O}:\mathcal{H}_\ell \to \mathcal{H}_\ell$ is an operator defined at the link $\ell$. Any state $\Psi_C(h_\ell)$ is a linear combination of product states of the form:
\be
\varphi_{\ell_1}(h_{\ell_1}) \cdots \varphi_{\ell_L}(h_{\ell_L}) \, .
\ee
The operator $\mathcal{O}_\ell$ is completely characterized by the action of $\mathcal{O}$ on the local states:
\[
\mathcal{O} \varphi_\ell(h_\ell) = \varphi_\ell'(h_\ell) \, .
\]

For any link $\ell$, we represent the local wavefunction for the link  $\ell^{-1}$ with the reverse orientation as
\[
\varphi_{\ell^{-1}}(h) = \varphi_\ell(h^{-1}) \, .
\]
Under an automorphism,
\begin{align*}
&U_A \mathcal{O}_\ell U_A^{-1} \left( \varphi_{\ell_1}(h_{\ell_1}) \cdots \varphi_{\ell_L}(h_{\ell_L}) \right) \\
&= U_A \mathcal{O}_\ell \left( \varphi_{\ell_1}(h_{A^{-1}({\ell_1})}) \cdots \varphi_{\ell_L}(h_{A^{-1}({\ell_L})}) \right)  \\
&= U_A \mathcal{O}_\ell \left( \varphi_{A(\ell_1)}(h_{\ell_1}) \cdots \varphi_{A(\ell_L)}(h_{\ell_L}) \right) \\
&= U_A \left( \varphi_{A(\ell_1)}(h_{\ell_1}) \cdots (\mathcal{O}\varphi_{A(\ell)})(h_\ell) \cdots \varphi_{A(\ell_L)}(h_{\ell_L}) \right) \\
&=\varphi_{\ell_1}(h_{\ell_1}) \cdots (\mathcal{O}\varphi_{A(\ell)})(h_{A(\ell)}) \cdots \varphi_{\ell_L}(h_{\ell_L}) \\
&=\mathcal{O}_{A(\ell)} \left( \varphi_{\ell_1}(h_{\ell_1}) \cdots \varphi_{\ell_L}(h_{\ell_L}) \right) \, .
\end{align*}
The local operator at the link $\ell$ was transformed by the automorphism into a local operator $\mathcal{O}_{A(\ell)}$ at the link $A(\ell)$:
\[
U_A \mathcal{O}_\ell U_A^{-1} = \mathcal{O}_{A(\ell)} \, .
\]
Let $\bar{\ell} \in \{\ell_i\}$ be the link such that $A(\ell)$ equals $\bar{\ell}$ or its inverse $\bar{\ell}^{-1}$. Then the action of $\mathcal{O}_{A(\ell)}$ on the local wavefunction at $\bar{\ell}$, $\varphi_{\bar{\ell}}(h_{\bar{\ell}}) \mapsto (\mathcal{O}\varphi_{A(\ell)})(h_{A(\ell)})$, is given by:
\[
(\mathcal{O}\varphi_{A(\ell)})(h_{A(\ell)}) = \begin{cases}
(\mathcal{O} \varphi_{\bar{\ell}})(h_{\bar{\ell}}) \, , & 
\text{if } A(\ell)= \bar{\ell} \, , \\
(\mathcal{O} \varphi_{\bar{\ell}^{-1}})(h_{\bar{\ell}}^{-1}) \, , & 
\text{if } A(\ell)= \bar{\ell}^{-1} \, .
\end{cases}
\]

An observable $\mathcal{O}^{1L}_{\mathrm{inv}}$ invariant under autormorphisms can be associated with a local one-link operator $\mathcal{O}_\ell$ through the application of the general prescription presented in Eq.~\eqref{eq:O-inv}:
\begin{align*}
\mathcal{O}^{1L}_{\mathrm{inv}} &= \frac{1}{|\mathrm{Aut }(\Gamma_H)|} \,\, \sum_A U_A \, \mathcal{O}_\ell \, U_A^{-1}  \\
&= \frac{1}{|\mathrm{Aut }(\Gamma_H)|} \,\, \sum_A \, \mathcal{O}_{A(\ell)} \\
&= \frac{1}{|\mathrm{Aut }(\Gamma_H)|} \,\, \sum_{\bar{\ell}} \, \left| [\mathrm{Aut }\; (\Gamma_H)]_{\ell\to \bar{\ell}} \right| \mathcal{O}_{\bar{\ell}} \, 
\end{align*}
where the sum in the last line runs over all oriented links $\bar{\ell}$. In analogy with the case of local node observables, we can simplify the expression by proving that the number $\left| [\mathrm{Aut }\; (\Gamma_H)]_{\ell\to \bar{\ell}} \right|$ is independent of $\bar{\ell}$. The argument is a simple adaptation of the previous case, as follows. 

Let $[\mathrm{Aut }\; (\Gamma_H)]_{\ell\to \bar{\ell}}=\{f_1, \dots, f_p\}$ be the set of automorphisms that map $\ell$ to $\bar{\ell}$, and $[\mathrm{Aut }\; (\Gamma_H)]_{\ell}=\{h_1,\dots, h_q\}$ be the set of automorphisms that preserve $\ell$. Any $f_i \in [\mathrm{Aut }\; (\Gamma_H)]_{\ell\to \bar{\ell}}$ can be written as the composition $f_1 \circ (f_1^{-1} \circ f_i)$ of an element $f_1^{-1} \circ f_i \in [\mathrm{Aut}\; (\Gamma_H)]_\ell$ and a fixed automorphism $f_1$ of $[\mathrm{Aut }\; (\Gamma_H)]_{\ell\to \bar{\ell}}$. If $f_i \neq f_j$, then $f_1^{-1} \circ f_i \neq f_1^{-1} \circ f_j$, since $f_1^{-1}$ is invertible.
Hence, $\left| [\mathrm{Aut }\; (\Gamma_H)]_{\ell\to \bar{\ell}} \right| \leq \left| [\mathrm{Aut}\; (\Gamma_H)]_\ell \right|$. Moreover, for any $h_i$, the composition $f_1 \circ h_i$ is in $[\mathrm{Aut }\; (\Gamma_H)]_{\ell\to \bar{\ell}}$. If $h_i \neq h_j$, then $f_1 \circ h_i \neq f_1 \circ h_j$, since $f_1$ is invertible. Therefore, $\left| [\mathrm{Aut}\; (\Gamma_H)]_\ell \right| \leq \left| [\mathrm{Aut }\; (\Gamma_H)]_{\ell\to \bar{\ell}} \right|$, and the result follows.

From the fact that the quantities $\left| [\mathrm{Aut }\; (\Gamma_H)]_{\ell\to \bar{\ell}} \right|$ are independent of $\bar{\ell}$ and sum to $|\mathrm{Aut }(\Gamma_H)|$, we obtain:
\[
\mathcal{O}^{1L}_{\mathrm{inv}} = \frac{1}{2L} \,\, \sum_{\bar{\ell}} \,  \mathcal{O}_{\bar{\ell}} \, ,
\]
where the sum runs over all oriented links, including two terms for each unoriented link. For an observable such that $\mathcal{O}_{\bar{\ell}}=\mathcal{O}_{\bar{\ell}^{-1}}$, i.e., that is invariant under link reversals, we can write 
\be
\mathcal{O}^{1L}_{\mathrm{inv}} = \frac{1}{L} \,\, \sum_{\ell} \,  \mathcal{O}_{\ell} \, ,
\label{eq:inv-one-link-obs}
\ee
where the sum runs over all links, each with a fixed, arbitrary orientation. We call an observable of the form \eqref{eq:inv-one-link-obs} an invariant one-link observable.

An example of an invariant one-link observable is the area $\hat{\mathcal{A}}_C$ at a link
\be
\hat{\mathcal{A}}_C = \frac{1}{L} \, \sum_\ell \mathds{1} \otimes \cdots  \otimes \hat{\mathcal{A}}_\ell \otimes \cdots \otimes \mathds{1} \, .
\ee
Two-link operators and higher can be constructed in a straightforward way.

\begin{figure}
\hspace{1cm}
\includegraphics[scale=0.4]{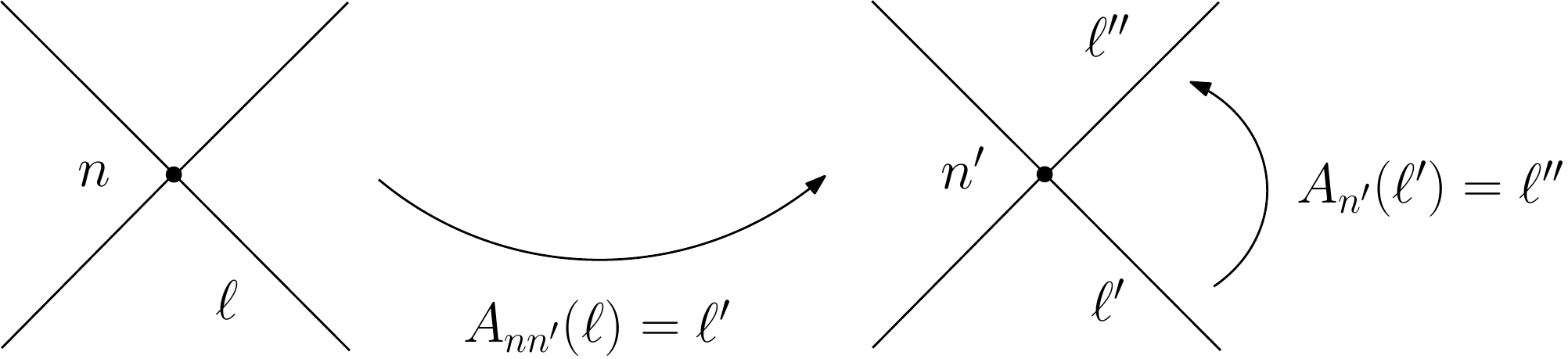}
\caption{Action of automorphisms on a $2$-CH graph. The node $n$ can be mapped into any node $n'$ by some automorphism $A_{nn'}$. The link $\ell$ is mapped into the link $\ell'$. Any two links $\ell'$ and $\ell''$ at a node are related by an automorphism $A_{n'}$.}
\label{fig:2chgraph-states}
\end{figure}

The invariant observables associated with $\mathcal{O}_\ell$ and its image $\mathcal{O}_{A(\ell)}$ under some automorphism are the same. It follows then from Eq.~\eqref{eq:invariant-exp} that, for states in $\mathcal{K}_{\Gamma_C}$, the averages of $\mathcal{O}_\ell$ and $\mathcal{O}_{A(\ell)}$ are the same. In particular, for any pair of links $na,nb$ at a node $n$, the averages of local link operators $\mathcal{O}_{na}$ and $\mathcal{O}_{nb}$ must be the same, and as a result independent of the direction from the node $n$ in the graph. This constitutes a discrete version of the property of isotropy: the statistics of local quantities measured at a single link is independent of the choice of the link at any node of the graph.

\subsubsection{$k$-CH graphs}
\label{sec:k-CH}

It was shown in Section \ref{sec:1-ch} that the reduced density matrix $\rho_{\{j_{ra}\}}$ for a single node with boundary spins $\{j_{ra}\}$ is the same for all nodes on a $1$-CH graph $\Gamma_H$, that is, $\rho_{\{j_{ra}\}}=\rho_{\{j_{sa}\}}$, $\forall r,s \in N(\Gamma_H)$. This result can be extended to connected regions formed by $k$ nodes in a $k$-CH graph. We shall now prove this generalization.

Let $\Gamma$ be a $k$-CH graph, and $\Gamma_R, \Gamma_{R'} \subset \Gamma$ be isomorphic connected subgraphs of $\Gamma$ with $k$ nodes. Then there is an automorphism $A \in \mathrm{Aut}(\Gamma)$ that maps $\Gamma_R$ into $\Gamma_{R'}$, due to the $k$-CH property. Let $\{j_{Ba}\}$ be the spins at links $\ell=Ba$ that connect a node in $N(\Gamma_R)$ to a node in its complement $\overline{N(\Gamma_R)}$. This set of links provides a natural notion of boundary for the region $R$ formed by the quantum polyhedra dual to nodes in $\Gamma_R$. We denote by $\{j_{Rb}\}$ the spins at links $\ell=Rb$ in $\Gamma_R$, and by $\{j_{\overline{R}c}\}$ the spins at links $\ell=\overline{R}c$ in the induced subgraph $\Gamma_{\overline{R}}$ with node set $\overline{N(\Gamma_R)}$, which represents the complement $\overline{R}$ of the region $R$. If we fix the boundary spins $\{j_{Ba}\}$, we obtain a subspace $\mathcal{H}_{\Gamma,\{j_{Ba}\}}$ that is a tensor product
\[
\mathcal{H}_{\Gamma,\{j_{Ba}\}} = \mathcal{H}_{\Gamma_R} \otimes \mathcal{H}_{\Gamma_{\overline{R}}}
\]
of an internal Hilbert space
\[
\mathcal{H}_{\Gamma_R} = \bigoplus_{\{j_{Rb}\}} \, \bigotimes_{n \in N(\Gamma_R)} \mathcal{H}_n \, ,
\]
and an external Hilbert space
\[
\mathcal{H}_{\Gamma_{\overline{R}}} = \bigoplus_{\{j_{\overline{R}c}\}} \, \bigotimes_{\overline{n} \in \overline{N(\Gamma_R)}} \mathcal{H}_{\overline{n}} \, ,
\]
with the boundary spins $\{j_{Ba}\}$ kept fixed in both cases. Let $P_{\{j_{Ba}\}}$ be the orthogonal projection to the subspace of fixed boundary spins, $P_{\{j_{Ba}\}} \mathcal{H}_{\Gamma}=\mathcal{H}_{\Gamma,\{j_{Ba}\}}$. For any boundary configuration, the reduced density matrix associated with $\Gamma_R$ is defined as:
\[
\rho_{\Gamma_R,\{j_{Ba}\}} = \Tr_{\mathcal{H}_{\Gamma_{\overline{R}}}} \left( P_{\{j_{Ba}\}} \ket{\Psi} \bra{\Psi} P_{\{j_{Ba}\}} \right) \, .
\]
We represent the state $\ket{\Psi}$ as in Eqs.~\eqref{eq:automorphism-spin-network-basis-1} and \eqref{eq:automorphism-spin-network-basis}, and automorphism-invariance again implies that $c'_{j'_{\ell}, i'_n}=c_{j_{\ell}, i_n}$, for $j'_\ell = j_{A^{-1} \ell} \, , i'_n = i_{A^{-1}n}$. It is convenient to explicitly distinguish the spins and intertwiners according to their position relative to the region $R$,
\[
c_{j_\ell,i_n} = c_{\{j_{Ba}\},\{j_{Rb}\},\{j_{\overline{R}c}\},\{i_n\},\{ i_{\overline{n}} \} } \, .
\]

We wish to compare the reduced density matrices for the subgraphs $\Gamma_R$ and $\Gamma_{R'}$. Let the boundary links of $\Gamma_{R'}$ be represented as $B'a=A(Ba)$. We denote the links of $\Gamma_{R'}$ by $R'b=A(Rb)$ and the links in the induced graph $\Gamma_{\overline{R'}}$ with node set $\overline{N(\Gamma_{R'})}$ by $\overline{R}'c=A(\overline{R}c)$. Then the reduced density matrix associated with $\Gamma_{R'}$ for boundary spins $\{j'_{B'a}\}$ reads:
\[
\rho_{\Gamma_{R'},\{j'_{B'a}\}} = \Tr_{\mathcal{H}_{\Gamma_{\overline{R'}}}} \left( P_{\{j'_{B'a}\}} \ket{\Psi} \bra{\Psi} P_{\{j'_{B'a}\}} \right) \, .
\]
Its matrix elements read:
\begin{align*}
&\bra{\{j'_{R'b}\},\kappa_1, \dots, \kappa_k}\rho_{\Gamma_{R'},\{j'_{B'a}\}} \ket{\{\tilde{j}'_{R'b}\},\tilde{\kappa}_1, \dots, \tilde{\kappa}_k} \\
&= \sum_{\{j'_{\overline{R}'c}\}} \sum_{i'_{\overline{n}'}} c'_{\{j'_{B'a}\},\{j'_{R'b}\},\{j'_{\overline{R}'c}\},\{\kappa_1, \dots, \kappa_k\},\{ i'_{\overline{n}'} \} } c'^*_{\{j'_{B'a}\},\{\tilde{j}'_{R'b}\},\{j'_{\overline{R}'c}\},\{\tilde{\kappa}_1, \dots, \tilde{\kappa}_k\},\{ i'_{\overline{n}'} \} } \\
&= \sum_{\{j_{\overline{R}c}\}} \sum_{i_{\overline{n}}} c_{\{j_{Ba}\},\{j_{Rb}\},\{j_{\overline{R}c}\},\{\kappa_1, \dots, \kappa_k\},\{ i_{\overline{n}} \} } c^*_{\{j_{Ba}\},\{\tilde{j}_{Rb}\},\{j_{\overline{R}c}\},\{\tilde{\kappa}_1, \dots, \tilde{\kappa}_k\},\{ i_{\overline{n}} \} } \\
&=\bra{\{j_{Rb}\},\kappa_1, \dots, \kappa_k}\rho_{\Gamma_{R},\{j_{Ba}\}} \ket{\{\tilde{j}_{Rb}\},\tilde{\kappa}_1, \dots, \tilde{\kappa}_k} \, ,
\end{align*}
where $j'_{R'b}=j_{Rb}$, $\tilde{j}'_{R'b}=\tilde{j}_{Rb}$, and the indices $\kappa_i,\tilde{\kappa}_i$ refer to the intertwiner bases $\{i'_n\}$ of nodes in $\Gamma'$ in the first line and to the intertwiner bases $\{i_n\}$ of nodes in $\Gamma$ in the last line. This shows that the matrix elements of the reduced density matrix are the same for the regions $R$ and $R'$.

\subsection{A cosmological sector}

On $2$-CH graphs, automorphism-invariant states were shown to display a discrete version of the properties of homogeneity and isotropy: the nodes that constitute the building blocks of space are indistinguishable, as well as directions from each node, represented by links. More general homogeneous and isotropic states can be built as superpositions of such states on distinct graphs. Let us introduce the Hilbert space
\be
\mathcal{K} = \bigoplus_{\Gamma_C} \mathcal{K}_{\Gamma_C} \, , 
\ee
which we call the space of cosmological states, where the direct sum runs over all $2$-CH graphs. States of the geometry in $\mathcal{K}$ are generic superpositions of homogeneous and isotropic quantum geometries over $2$-CH graphs,
\be
\ket{\Psi} = \sum_{\Gamma_\alpha} \ket{\Gamma_\alpha,\psi_\alpha} \, ,
\label{eq:generic-C-state}
\ee
where states on distinct graphs are orthogonal. For a normalized state, $\scalar{\Psi}{\Psi}=1$, the probability $p_\alpha$ associated with a given graph $\Gamma_\alpha$ is
\be
p_\alpha = \scalar{\Gamma_\alpha,\psi_\alpha}{\Gamma_\alpha,\psi_\alpha} \, .
\ee

Explicit examples of cosmological states in $\mathcal{K}$ are given by arbitrary superpositions of Bell-network states $\ket{\Gamma_C,\lambda,\mathcal{B}}$ on $2$-CH graphs $\Gamma_C$. As shown in Section \ref{sec:Bell-states}, Bell-network states are automorphism-invariant on any graph $\Gamma$; in particular, they are invariant on $2$-CH graphs. The projection of a state $\ket{\Gamma_C,\lambda,\mathcal{B}}$ to the subspace with fixed spins $j_\ell=j$ is also invariant, since these subspaces are invariant under the action of the automorphisms. Therefore, superpositions of Bell-network states $\ket{\Gamma_C, \{j_\ell=j\},\mathcal{B}}$ with identical fixed spins $j_\ell=j$ (see definition in Eq.~\eqref{BNdef}) also constitute simpler examples of cosmological states.

Graph-preserving operators on $\mathcal{K}$ can be defined by taking direct sums of operators defined on fixed graphs. Consider, for instance, a family of invariant one-node observables $\mathcal{O}_{\Gamma_\alpha}:\mathcal{K}_{\Gamma_\alpha} \to \mathcal{K}_{\Gamma_\alpha}$, each defined on a fixed $2$-CH graph $\Gamma_\alpha$. An observable on $\mathcal{K}$ can be defined through
\be
\mathcal{O}_C = \bigoplus_{\Gamma_\alpha} \mathcal{O}_{\Gamma_\alpha} \, .
\label{eq:O_C-def}
\ee
Its action on a generic state \eqref{eq:generic-C-state} reads
\begin{align*}
	\mathcal{O}_C \ket{\Psi} &= \mathcal{O}_C \sum_{\Gamma_\alpha} \ket{\Gamma_\alpha,\psi_\alpha} \\
	&= \sum_{\Gamma_\alpha} \mathcal{O}_{\Gamma_\alpha} \ket{\Gamma_\alpha,\psi_\alpha} \, ,
\end{align*}
and its expected value is
\begin{align}
\bra{\Psi} \mathcal{O}_C \ket{\Psi} &= \sum_{\Gamma_\alpha} p_\alpha \mean{\mathcal{O}_{\Gamma_\alpha}} \, , \nonumber \\
\mean{\mathcal{O}_{\Gamma_\alpha}} &= \frac{\bra{\Gamma_\alpha,\psi_\alpha} \mathcal{O}_{\Gamma_\alpha} \ket{\Gamma_\alpha,\psi_\alpha}}{\scalar{\Gamma_\alpha,\psi_\alpha}{\Gamma_\alpha,\psi_\alpha}} \, ,
\label{eq:O_C-average-general}
\end{align}
the average of the expected values of the observables over all graphs.

Of special interest are graph-preserving operators $\mathcal{O}_C$ with restrictions $\mathcal{O}_{\Gamma_\alpha}$ that describe a measurement of the same local quantity in all graphs. Consider, for instance, a measurement of the volume of a single node for a region with $V$ boundary links. Let $V_C$ be the operator on $\mathcal{K}$ that describes this measurement. The invariant volume operator at a node, for a fixed $1$-CH graph, was defined in Eqs.~\eqref{eq:volume-operator-inv} and \eqref{eq:volume-operator-node}. For each graph $\Gamma_\alpha$, it reads:
\be
V_\alpha = \frac{1}{N_\alpha} \sum_{n \in \Gamma_\alpha}  \mathds{1} \otimes \cdots  \otimes V_n \otimes \cdots \otimes \mathds{1} \,,
\ee
where $N_\alpha$ is the number of nodes in $\Gamma_\alpha$. The associated operator on $\mathcal{K}$ is
\be
V_C = \bigoplus_{\Gamma_\alpha} V_\alpha \, ,
\ee
where the sum runs over all $V$-valent $2$-CH graphs. Similarly, the total volume operator is defined on each graph by $V_{\alpha,tot}=N_\alpha V_\alpha$, and the associated operator on $\mathcal{K}$ is $V_{C,tot} = \bigoplus_{\Gamma_\alpha} V_{\alpha,tot}$.

Let us restrict to the case of graph-preserving operators $\mathcal{O}_C$ for which the operators $\mathcal{O}_{\Gamma_\alpha}$ are invariant one-node observables describing the same physical quantity (e.g., the volume of a node). By this we mean that all operators $\mathcal{O}_{\Gamma_\alpha}$ are associated with the same node observable $\mathcal{O}^1_\alpha$ on $\mathcal{H}^1$:
\be
\mathcal{O}^1_\alpha=\mathcal{O}^1 = \bigoplus_{\{j_a\}} \mathcal{O}_{\{j_a\}} \, ,
\label{eq:same-inv-operator}
\ee
where $\mathcal{H}^1$ is the abstract intertwiner space defined in Eq.~\eqref{eq:abstract-intertwiner-space}, and the node observable $\mathcal{O}^1_\alpha$ is defined in Eqs.~\eqref{eq:abstract-one-node-obs} and \eqref{eq:abstract-one-node-obs-2}. We will further discuss this condition later in this section, but for now let us assume that it is satisfied. We call such operators $\mathcal{O}_C$ invariant one-node observables on $\mathcal{K}$. From Eq.~\eqref{eq:abstract-one-node-av}, the mean valued $\mean{\mathcal{O}_{\Gamma_\alpha}}$ of its restriction to each graph is then given by
\[
\mean{\mathcal{O}_{\Gamma_\alpha}} = \Tr_{\mathcal{H}^1} (\rho^1_\alpha \mathcal{O}^1_\alpha) \, ,
\]
where the density matrix $\rho^1_\alpha$ is defined in Eq.~\eqref{eq:abstract-one-node-state}. It follows that
\be
\bra{\Psi} \mathcal{O}_C \ket{\Psi} = \Tr_{\mathcal{H}^1} \left( \rho_C  \, \mathcal{O}^1 \right) \, ,
\label{eq:O_C-average}
\ee
where
\be
\rho_C = \sum_{\Gamma_\alpha} p_\alpha \rho^1_\alpha \, ,
\label{eq:rho-C-def}
\ee
and the statistics of invariant one-node observables on $\mathcal{K}$ is completely described by a density matrix $\rho_C$ consisting of a mixture of the one-node density matrices for the individual graphs in superposition.

From Eq.~\eqref{eq:abstract-one-node-state}, we can express this density matrix as
\be
\rho_C = \sum_{\Gamma_\alpha} \sum_{\{j_a\}}p_\alpha p_{\alpha,\{j_a\}} \rho^1_{\alpha,\{j_a\}} \, .
\label{eq:rho-C}
\ee
Summing over the graphs, we obtain the representation
\be
\rho_C = \sum_{\{j_a\}} P_{\{j_a\}}\rho_{\{j_a\}} \, ,
\label{eq:rho-C-2}
\ee
with
\begin{align}
P_{\{j_a\}} &= \sum_{\Gamma_\alpha} p_\alpha p_{\alpha,\{j_a\}} \, , \nonumber \\  \rho_{\{j_a\}} &= \frac{1}{P_{\{j_a\}}} \sum_{\Gamma_\alpha} p_\alpha p_{\alpha,\{j_a\}} \rho^1_{\alpha,\{j_a\}} \, .
\end{align}
The density matrix $\rho_{\{j_a\}}$ encodes the statistics of measurements of the local observable at a node, given that spins $\{j_a\}$ where observed at its boundary, and $P_{\{j_a\}}$ is the probability of the configuration $\{j_a\}$. The graph and node at which the observation is performed remain unspecified: the invariant observable includes the possibility that the measurement is performed in any node of any graph in the superposition, and these possibilities are averaged over, for each boundary spin configuration.

Let us now discuss the interpretation of the condition \eqref{eq:same-inv-operator} that defines invariant one-node observables on $\mathcal{K}$. For that, suppose that a local property $\mathcal{O}$ of the geometry is measured at a node with $V$ links, say a dihedral angle, a face area, the volume, or any other local quantity, for a state that involves a superposition of states on distinct graphs. Any such a measurement must be represented by an invariant operator $\mathcal{O}_C:\mathcal{K} \to \mathcal{K}$ of the form \eqref{eq:O_C-def} that vanishes on graphs that are not $V$-valent. Choose a reference node $r_\alpha$ with links $r_{\alpha}a$ in each $V$-valent graph. Now let $\mathcal{I}$ be a local quantity represented by an observable
\[
\mathcal{I}_C = \bigoplus_{\Gamma_\alpha} \mathcal{I}_{\Gamma_\alpha} \, ,
\]
where the sum runs over all $V$-valent $2$-CH graphs, and the operators $\mathcal{I}_{\Gamma_\alpha}:\mathcal{K}_{\Gamma_\alpha} \to \mathcal{K}_{\Gamma_\alpha}$ are invariant one-node observables on $\mathcal{K}_{\Gamma_\alpha}$,
\[
\mathcal{I}_{\Gamma_\alpha} = \frac{1}{N_\alpha} \sum_{n_\alpha \in \Gamma_\alpha} \mathcal{I}_{n_\alpha} \, ,
\]

Suppose that, for any configuration of boundary spins $j_{r_\alpha a}$, the restriction $\mathcal{I}_{\{j_{r_\alpha a}\}}$ of the local operator $\mathcal{I}_{r_\alpha}$ to the intertwiner space for such spins has eigenvalues $\mathcal{I}_\kappa$ that are all distinct, where $\kappa=1,\dots,d$, with $d=\dim \mathcal{H}_{r_\alpha}$. For instance, in a four-valent node, $\mathcal{I}$ can be the spin at a virtual link connecting the links $1$ and $2$ to the links $3$ and $4$ of the node. Let $\ket{r_{\alpha},\kappa} \in \mathcal{H}_{r_{\alpha}}$ be the normalized eigenstate associated with the eigenvalue $\mathcal{I}_\kappa$. Then the sets $\{\ket{r_{\alpha},\kappa}; \kappa=1,\dots, d\}$ provide orthonormal bases for the intertwiner spaces at each node $r_{\alpha}$. By identifying states $\ket{r_{\alpha},\kappa} \simeq \ket{r_{\beta},\kappa}$, we obtain isomorphisms $G_{\alpha \beta}:\mathcal{H}_{r_{\alpha}} \to \mathcal{H}_{r_{\beta}}$ with action
\be
G_{\alpha \beta}\ket{r_{\alpha},\kappa} = \ket{r_{\beta},\kappa} \, .
\label{eq:G-alpha-beta}
\ee
 
Similarly, local operators $\mathcal{O}_{n_\alpha}$ are associated with the quantity $\mathcal{O}$. Denote by $\mathcal{O}_{\{j_{r_\alpha a}\}}$ the restriction of the local operator $\mathcal{O}_{r_\alpha}$ to the subspace with boundary spins $\{j_{r_\alpha a}\}$. If the corresponding operator $\mathcal{O}_{\{j_{r_\beta a}\}}$ in the graph $\Gamma_\beta$, for the same boundary spins $j_{r_\alpha a}=j_{r_\beta a}=j_a$, is given by the image of $\mathcal{O}_{\{j_{r_\alpha a}\}}$ under the isomorphism that maps $\mathcal{H}_{r_\alpha}$ into $\mathcal{H}_{r_\beta}$, $\mathcal{O}_{\{j_{r_\beta a}\}}=G_{\alpha \beta} \mathcal{O}_{\{j_{r_\alpha a}\}} G_{\alpha \beta}^{-1}$, then the operators $\mathcal{O}_{\{j_{r_\alpha a}\}}$ and $\mathcal{O}_{\{j_{r_\beta a}\}}$ have the same matrix elements in the bases $\{\ket{r_\alpha,\kappa}\}$ and $\{\ket{r_\beta,\kappa}\}$:
\be
\bra{r_\alpha,\kappa'} \mathcal{O}_{\{j_{r_\alpha a}\}} \ket{r_\alpha,\kappa''} = \bra{r_\beta,\kappa'} \mathcal{O}_{\{j_{r_\beta a}\}} \ket{r_\beta,\kappa''} \, .
\label{eq:O-change-basis}
\ee
We interpret this as representing that the same quantity was measured on all graphs.

We can now compute the node observables $\mathcal{O}^1_\alpha$ on the abstract intertwiner space $\mathcal{H}^1$, for each $\mathcal{O}_{\Gamma_\alpha}$, following the general prescription described in Section \ref{sec:1-ch}. As discussed there, the matrix elements of $\mathcal{O}^1_\alpha$ are independent of the choice of reference node in $\Gamma_\alpha$ if the intertwiners bases are chosen in a consistent way for all nodes in $\Gamma_\alpha$. In addition, for a superposition of graphs, we can choose intertwiner bases $\{\ket{i_{r_\alpha}}\}$ that satisfy
\be
\ket{i_{r_\beta}=\kappa}=G_{\alpha \beta} \ket{i_{r_\alpha}=\kappa} \, .
\label{eq:consistency-cond-2}
\ee
The matrix elements of $\mathcal{O}^1_\alpha$ can be computed in such bases by taking the nodes $r_\alpha$ as reference nodes on each graph, leading to:
\begin{align}
\bra{i_{r_\beta}=\kappa} \mathcal{O}_{\{j_{r_\beta a}\}} \ket{i_{r_\beta}=\tilde{\kappa}} &=  \sum_{\kappa'\kappa''} \langle i_{r_\beta}=\kappa | r_\beta,\kappa' \rangle \bra{r_\beta,\kappa'} \mathcal{O}_{\{j_{r_\beta a}\}} \ket{r_\beta,\kappa''} \langle r_\beta,\kappa'' | i_{r_\beta}=\tilde{\kappa} \rangle \nonumber \\
&=  \sum_{\kappa'\kappa''} \langle i_{r_\alpha}=\kappa | r_\alpha,\kappa' \rangle \bra{r_\alpha,\kappa'} \mathcal{O}_{\{j_{r_\alpha a}\}} \ket{r_\alpha,\kappa''} \langle r_\alpha,\kappa'' | i_{r_\alpha} =\tilde{\kappa} \rangle \nonumber \\
&= \bra{i_{r_\alpha}=\kappa} \mathcal{O}_{\{j_{r_\alpha a}\}} \ket{i_{r_\alpha}=\tilde{\kappa}} \, ,
\end{align}
where we used the Eqs.~\eqref{eq:G-alpha-beta}, \eqref{eq:O-change-basis} and \eqref{eq:consistency-cond-2}. The matrix elements of $\mathcal{O}_{\{j_{r_\beta a}\}}$ are then the same for all graphs, which implies that $\mathcal{O}^1_\alpha$ has the same form on all graphs. The condition \eqref{eq:same-inv-operator} is thus satisfied.

The construction of invariant one-node observables and the reduced density matrix for a single node, while considered here for the case of interest of homogeneous and isotropic spaces described by $2$-CH graphs, extends without modifications to a superposition of $1$-CH graphs, since only the $1$-CH property was used. In addition, for $2$-CH graphs, not only isolated nodes are all equivalent, but also pairs of adjacent nodes can always be related by some automorphism. As a result, the reduced density matrix of a pair of adjacent nodes for a given boundary geometry has always the same form on a $2$-CH graph, regardless of the choice of the pair of adjacent nodes, as discussed in Section \ref{sec:k-CH}. One can then introduce an invariant two-node reduced density matrix that describes the statistics of invariant observables representing measurements performed on a pair of adjacent nodes, for any superposition of states on $2$-CH graphs, in analogy with the case of a single node. More generally, in $k$-CH graphs, isomorphic subgraphs with $k$ nodes can be related by automorphisms, and an invariant reduced density matrix can be assigned to such regions.


\section{Entanglement entropy for a superposition of graphs}
\label{sec:entropy}

In the last section, we introduced the density matrix $\rho_C \in \mathcal{H}^1$ that describes the statistics of invariant one-node observables for a state involving a superposition of graphs. This density matrix is the restriction of a state $\ket{\Psi} \in \mathcal{K}$ to the subalgebra of observables formed by invariant one-node observables, which characterizes a subsystem composed of a single node in an automorphism-invariant way.

The density matrix $\rho_C$ provides us with a tool to introduce a notion of geometric entropy in $\mathcal{K}$ for a subsystem formed by a single node. The entropy of a node is defined as the von Neumann entropy associated with the reduced density matrix $\rho_C$:
\be
S_C = - \Tr_{\mathcal{H}^1} ( \rho_C \log \rho_C) \, .
\label{eq:see-one-node}
\ee
The entropy is completely determined by $\rho_C$, and therefore invariant under automorphisms and well defined for states involving superpositions of graphs. Let us express it in a more explicit form in order to discuss its main properties.

From Eq.~\eqref{eq:rho-C-2}, the density matrix $\rho_C$ has a block diagonal form with respect to the direct sum $\mathcal{H}^1 = \bigoplus_{\{j_a\}} \mathcal{H}_{\{j_a\}}$. The trace can then be decomposed into a sum of contributions from each component, and we obtain the formula:
\be
S_C = - \sum_{\{j_a\}} P_{\{j_a\}} \log P_{\{j_a\}} + \sum_{\{j_a\}} P_{\{j_a\}} S_{\{j_a\}} \, ,
\label{eq:total-entropy}
\ee
where
\[
S_{\{j_a\}} = - \Tr_{\mathcal{H}_{\{j_a\}}} \left( \rho_{\{j_a\}} \log \rho_{\{j_a\}} \right)
\]
is the entropy for fixed boundary spins ${\{j_a\}}$. The entropy $S_C$ includes a contribution from the classical distribution of probabilities $P_{\{j_a\}}$ for the spins, given by the first term in Eq.~\eqref{eq:total-entropy}, and independent contributions from each spin configuration ${\{j_a\}}$, given by the von Neumann entropy of the associated density matrix $\rho_{\{j_a\}}$, weighted by the spin configuration probabilities.

When the state of the geometry does not involve a superposition of graphs, the probability $P_{\{j_a\}}$ and the reduced density matrix $\rho_{\{j_a\}}$ reduce to that of any single node in the graph. The formula \eqref{eq:total-entropy} then corresponds to that of a mixture of the reduced density matrices for all spin configurations, at any node, weighted by the associated probabilities, as could be expected. Such a formula for the entropy was previously employed for the analytical and numerical determination of the entropy of Bell states on fixed graphs in \cite{bell-entropy}, where only CH graphs were considered. The entropy formula \eqref{eq:total-entropy} obtained in our invariant approach thus reduces to that considered in \cite{bell-entropy} at a fixed graph and extends it to the case of a superposition of graphs.

The distribution of probabilities $p_i$ for a superposition of graphs $\Gamma_i$ contributes to the entanglement entropy differently from the probability distribution for the spin configurations. Consider, for example, a state for which the one-node density matrices $\rho^{1}_i$ are the same for all graphs $\Gamma_i$, that is, $\rho^{1}_i=\rho^{1}$. From \eqref{eq:rho-C-def}, it follows that $\rho_C=\rho^{1}$. The entropy $S_C$ is then the same as for any individual graph in the superposition. That this must be so follows from the fact that the entropy $S_C$ is completely determined by the statistics of observations performed within a node, given that some spin configuration was observed at its boundary. If the geometries of the nodes are the same, for any boundary configuration, for all graphs in the superposition, then the entropy should indeed not be affected by the graph superposition.

As discussed in the previous section, for cosmological states based on $2$-CH graphs, a reduced density matrix can also be associated in an invariant way with regions formed by two adjacent nodes. In analogy with the case of a single node, the entropy of a region formed by two adjacent nodes can then be defined in an automorphism-invariant way as the von Neumann entropy of such a two-node reduced density matrix. More generally, the entropy of regions formed by $k$ nodes can be analogously defined for states formed by the superposition of automorphism-invariant states on $k$-CH graphs.

We will now illustrate the application of the entropy formula \eqref{eq:total-entropy} through the analysis of a concrete example of invariant state involving a superposition of graphs.

\subsection{Superposition of Bell states on $\Gamma_5$ and $\Gamma_6$}

Let $\Psi \in \mathcal{K}$ be a superposition
\begin{align}
\ket{\Psi} = \sqrt{p} \ket{\Gamma_5, \psi_5} + \sqrt{1-p} \ket{\Gamma_6,\psi_6}
\label{eq:superposition56}
\end{align}
of a state $\ket{\Gamma_5, \psi_5}$ on the pentagram $\Gamma_5$ (Fig.~\ref{fig:4-regular-homogeneous}(a)) and a state $\ket{\Gamma_6,\psi_6}$ on the T\`uran graph $\mathcal{T}(6,3)$ (Fig.~\ref{fig:4-regular-homogeneous}(b)), where $p\in[0,1]$ is the probability of the graph $\Gamma_5$. Both graphs are four-regular and homogeneous. From their homogeneity, they must also be 2-CH graphs. We take $\ket{\Gamma_5, \psi_5}$ and $\ket{\Gamma_6,\psi_6}$ to be Bell-network states with fixed spins $j_\ell=1/2$ (Eq.~\eqref{BNdef}),
\begin{align*}
\ket{\Gamma_5, \psi_5} &= \ket{\Gamma_5, \{j_\ell=1/2\},\mathcal{B}}  \\
	&= \sqrt{\frac{18}{7}} \sum_{i_n=0}^1 \, \overline{\mathcal{A}_{\Gamma_5} \left(j_\ell, i_n \right)} \,\, \ket{\Gamma_5, \{j_\ell=1/2\}, \{i_n\}} \, , \\
\ket{\Gamma_6,\psi_6}  &= \ket{\Gamma_6, \{j_\ell=1/2\},\mathcal{B}} \\
	&= \sqrt{\frac{324}{73}} \sum_{i_n=0}^1 \, \overline{\mathcal{A}_{\Gamma_6} \left(j_\ell, i_n \right)} \,\, \ket{\Gamma_6, \{j_\ell=1/2\}, \{i_n\}} \, ,
\end{align*}
The sum over intertwiners restricts to $i_n=0,1$ because all links are colored with spins $j_\ell=1/2$. At each node, we choose the intertwiner basis:
\begin{align*}
[i_0]^{m_1 m_2 m_3 m_4} &= \frac{1}{2} \epsilon_{m_1 m_2} \epsilon_{m_3 m_4} \, , \\
[i_1]^{m_1 m_2 m_3 m_4} &= \frac{1}{2\sqrt{3}} \sum_{i=1}^3 (-1)^{-m_2 -m_4} [\sigma_1]^{m_1}_{-m_2} [\sigma_1]^{m_3}_{-m_4} \, ,
\end{align*}
where
\[
\epsilon_{mm'} = \begin{pmatrix}
		0 & 1 \\ -1 & 0
		\end{pmatrix} \, ,
\]
and the matrices $[\sigma_i]^m_{m'}$ are the Pauli matrices, with $m,m'=\pm 1/2$.

We denote by $\rho^1_5$ and $\rho^1_6$ the invariant one-node density matrices of $\ket{\Gamma_5, \psi_5}$ and $\ket{\Gamma_6, \psi_6}$, respectively. In order to determine them, we compute the reduced density matrix of the corresponding states at an arbitrary node on each graph. We find:
\begin{align*}
\rho_5^1 &= \frac{1}{2} (\ket{0}\bra{0}+\ket{1}\bra{1}) \, , \\
\rho_6^1 &= \frac{1}{73} \left( 34 \, \ket{0}\bra{0} - \frac{5\sqrt{3}}{2} (\ket{0}\bra{1}+\ket{1}\bra{0}) + 39 \, \ket{1}\bra{1} \right) \, .
\end{align*}
We checked that the same density matrix is obtained for all nodes in each graph.  The state $\rho_5^1$ is maximally mixed, with entropy $S^1_5=\ln 2 \simeq 0.6931$. The entropy $S^1_6$ of $\rho_6^1$ is slightly smaller, $S^1_6 \simeq 0.6837$.

\begin{figure}
\includegraphics[scale=0.85]{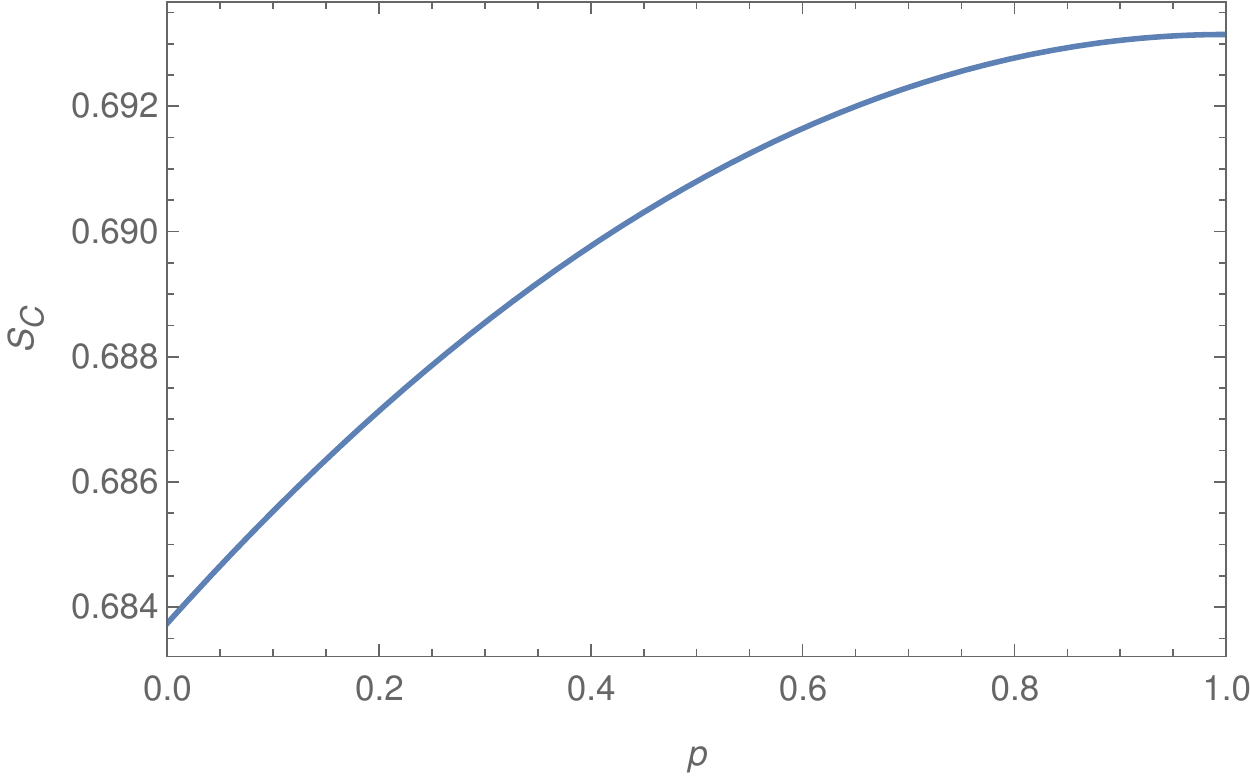}
\caption{Entropy $S_C$ of a single node for a superposition of Bell-network states on graphs $\Gamma_5$ and $\Gamma_6$ described by the state~\eqref{eq:superposition56} as a function of the probability $p$ associated with the graph $\Gamma_5$.}
\label{fig:entropy1-superposition}
\end{figure}

The invariant one-node density $\rho_C$ for the state $\ket{\Psi}$ obtained by the superposition of the states on both graphs is given by Eq.~\eqref{eq:rho-C-def}. We obtain:
\[
\rho_C = p \rho_5^1 + (1-p) \rho_6^1 \, .
\]
The entropy of the node is given by Eq.~\eqref{eq:total-entropy}. Its dependence on the probability $p$ of the graph $\Gamma_5$ is plotted in Fig.~\ref{fig:entropy1-superposition}. As the probability $p$ varies from $0$ to $1$, the entropy interpolates between the values of the entropies of the individual states $\ket{\Gamma_5, \psi_5}$ and $\ket{\Gamma_6, \psi_6}$ of the superposition, and is maximal for $p=1$, when the superposition becomes trivial and the state of the geometry reduces to $\ket{\Gamma_5, \psi_5}$.

\begin{figure}[htbp]
\includegraphics[scale=0.85]{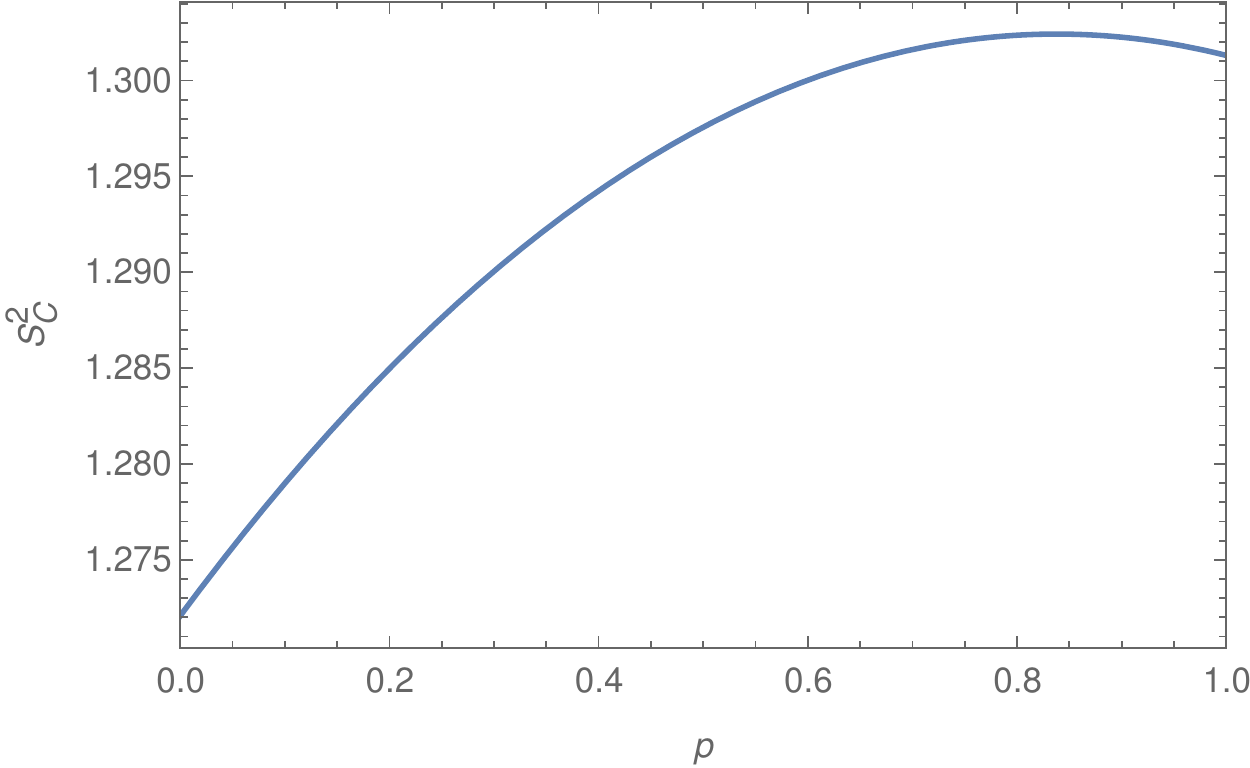}
\caption{Entropy $S^2_C$ of two adjacent nodes for a superposition of Bell-network states on graphs $\Gamma_5$ and $\Gamma_6$ described by the state~\eqref{eq:superposition56} as a function of the probability $p$ associated with the graph $\Gamma_5$.}
\label{fig:entropy2-superposition}
\end{figure}

We can also compute the entropy of a region formed by two adjacent nodes. We denote by $\rho^2_5$ and $\rho^2_6$ the invariant density matrices for two adjacent nodes of $\ket{\Gamma_5, \psi_5}$ and $\ket{\Gamma_6, \psi_6}$, respectively. As for a single node, we can compute the reduced density matrix of each invariant state for any pair of adjacent nodes. Since both graphs are $2$-CH, the result is independent of the choice of adjacent nodes. We find:
\begin{gather*}
\rho_5^2 = \frac{1}{28} \begin{pmatrix} 
		6 &  - \sqrt{3} & - \sqrt{3} & 0 \\
		- \sqrt{3} & 8 & 2 & \sqrt{3} \\
		- \sqrt{3} & 2 & 8 & \sqrt{3} \\
		0 & \sqrt{3} & \sqrt{3} & 6
				\end{pmatrix} \, , \\
\rho_6^2 = \frac{1}{292} \begin{pmatrix} 
		63 & -15\sqrt{3} & -15\sqrt{3} & -3 \\
		-15\sqrt{3} & 93 & 3 & 5\sqrt{3} \\
		-15\sqrt{3} & 3 & 93 & 5\sqrt{3} \\
		-3 & 5\sqrt{3} & 5\sqrt{3} & 43
				\end{pmatrix}\, ,
\end{gather*}
in the basis $\{\ket{00}, \ket{01}, \ket{10}\, \ket{11}\}$. The entropy of the state $\rho_5^2$ is $S_5^2 \simeq 1.30131$, and that of the state $\rho_5^2$ is $S_6^2 \simeq 1.27207$. Again, the entropy is smaller in the graph $\Gamma_6$. For the superposition \eqref{eq:superposition56}, the two-node density matrix is
\[
\rho^2_C = p \rho_5^2 + (1-p) \rho_6^2 \, ,
\]
and the entropy $S_C^2$ of two adjacent nodes is the von Neumann entropy of $\rho^2_C$. Its dependence on the probability $p$ of the graph $\Gamma_5$ is plotted in Fig.~\ref{fig:entropy2-superposition}. The entropy has a maximum at $p \simeq 0.84$. Therefore, in this example the entropy of the superposition can exceed that of each graph in the superposition. One can expect the same to occur for states involving superpositions of a larger number of more complex graphs.


\section{Summary and discussion}
\label{sec:discussion}

We introduced a general space $\mathcal{K}$ of quantum states of the geometry in loop quantum gravity whose symmetry properties correspond to discrete versions of the properties of homogeneity and isotropy, and constructed a nontrivial family of concrete examples consisting of Bell-network states \cite{bellnetwork} on generic $2$-CH graphs. Such states constitute a new class of quantum geometries that can be used as boundary states for the computation of transition amplitudes in spinfoam cosmology, or as initial states for a Hamiltonian description of the dynamics of quantum cosmological spacetimes. The local geometry of a generic cosmological state $\ket{\Psi} \in \mathcal{K}$, which can include superpositions of states defined on distinct graphs, is captured by one-node observables analogous to the one-body operators routinely used in many-body quantum mechanics \cite{martin-rothen}. We derived an explicit formula for the density matrix describing the restriction of a cosmological state $\ket{\Psi}$ to the algebra of one-node observables. The von Neumann entropy of this density matrix provides a natural definition of the entanglement entropy associated with a node. We computed the geometric entanglement entropy for a superposition of Bell states on distinct graphs in order to illustrate the application of the techniques introduced here in an explicit example.

Homogeneous and isotropic states were first defined on fixed $2$-CH graphs $\Gamma_C$, a class of highly symmetric graphs that includes dual graphs of regular discretizations of homogeneous and isotropic spaces, and generalizes, in particular, cubulations and regular decompositions of the $3$-sphere. The automorphism group of a $2$-CH graph is node- and link-transitive, which implies that all its nodes are equivalent, as well as all its links. We adopted the requirement that physical states and observables of the geometry must be invariant under all automorphisms of the graph \cite{rovellicombi}. The symmetries of the graphs are then inherited by the states of the geometry, and measurements of the geometry cannot distinguish among the nodes or the links of the graph. The indistinguishability of the nodes characterizes a notion of homogeneity for the quantum geometry, and that of the links at any node characterizes a discrete notion of isotropy. A generic cosmological state was then defined as a generic superposition of automorphism-invariant states on $2$-CH graphs.

We proved that a special class of Bell-network states satisfies the proposed conditions of homogeneity and isotropy. Bell-network states are constructed by using squeezed vacua techniques in the bosonic representation of LQG \cite{squeezedvacua,bosonic,bellnetwork}. 
In this representation, opposite endpoints of a link represent faces of adjacent polyhedra that are glued along the link. The building blocks of a Bell-network state are individual link states whose opposite endpoints are maximally entangled. In addition, each link state is invariant under orientation reversal. Gluing identical copies of such maximally entangled pairs according to the graph structure produces automorphism-invariant states $\ket{\Gamma_C, \lambda ,\mathcal{B}}$ on $2$-CH graphs $\Gamma_C$, as desired. In contrast with coherent states \cite{livine-speziale,coherent-1, coherent-2}, the construction of Bell-network states does not require the specification of a local classical discrete geometry at each node on which the state would be peaked on. Instead, they are superpositions of states weighted by their $SU(2)$ symbols and a simple function of the parameter $\lambda$, which on the average produce a regular geometry. On a pentagram, for instance, in the large spin limit, a Bell-network state at fixed spins is a superposition of coherent states peaked on all classical vector geometries with equal weights \cite{bellnetwork}. The homogeneity and isotropy of the state result from the uniform superposition of states peaked on a large family of classical geometries.

As the states $\ket{\Gamma_C, \lambda ,\mathcal{B}}$ are not peaked on a classical piecewise linear geometry, their effective geometry does not necessarily correspond to that of flat polyhedra. The mean volume and boundary area of a node, for instance, may not always be related as the volume and boundary area of a regular tetrahedron. As a result, the effective geometry might better approximate a regular curved tetrahedron \cite{vinberg}. We will analyze the dependence of the effective geometry of Bell-network states on the parameter $\lambda$ and graph $\Gamma_C$ in our further works in order to clarify whether they can be used to describe an effective geometry formed by gluing regular curved polyhedra.

The condition that states and observables of the geometry are automorphism-invariant severely restricts the space of states and the algebra of observables in the highly symmetric $2$-CH graphs. In particular, observables that act nontrivially only on a single node, as local area and volume operators, are not invariant. In order to obtain observables that describe measurements on a single polyhedron while respecting automophism invariance, we applied a group averaging to the local operators. On $2$-CH graphs, this produces invariant one-node observables that are proportional to sums of local operators over all nodes of the graph, analogous to one-body operators in many-body quantum mechanics. The invariant volume operator, for instance, is the average of the volume operators of all nodes of the lattice. Such one-node observables describe measurements performed on a single node, which remains unspecified. By taking their direct sum over all $2$-CH graphs, we constructed invariant observables that act on generic cosmological states, possibly including superpositions of states on distinct graphs. They describe measurements performed on a single node, whose graph and location on a graph remain unspecified. The resulting formalism is reminiscent of the framework of group field theory \cite{gft}, in which the quantum geometry is described in terms of excitations of indistinguishable building blocks of space, and cosmological degrees of freedom can be captured by one-body operators that describe averaged properties of the geometry \cite{gft-cosmology-3}. In our case, we arrived at the one-node observables as a direct consequence of the requirement of automorphism-invariance on $2$-CH graphs in the standard formalism of loop quantum gravity. We can further analyze the construction of cosmological states, which can be linked to group field theory formalism. 

The space of cosmological states does not have a natural decomposition into a tensor product of local spaces. In fact, there are no local states in $\mathcal{K}_{\Gamma_C}$, as any state on a $2$-CH graph is completely delocalized, due to its invariance under the automorphism group, which is node-transitive. In addition, the geometry can involve a superposition of states on distinct graphs. Nevertheless, we showed that a notion of entanglement entropy of a single node can be introduced. Suppose that a measurement of the spins at a node returns a spin configuration $\{j_a\}$ with a probability $P_{\{j_a\}}$. One can then ask about the statistics of volume measurements, for instance, or of any other internal property of the dual polyhedron, under the condition that the boundary spins $\{j_a\}$ were observed. The statistics of one-node observables describing such measurements is completely characterized by a reduced density matrix $\rho_C = \sum_{\{j_a\}} P_{\{j_a\}}\rho_{\{j_a\}} $. We defined the entanglement entropy of the node as the von Neumann entropy of $\rho_C$. In this way, instead of describing a local region by selecting a specific node in a graph, we characterized it by the measurement of a boundary geometry, which can be in any graph, and the entropy of the region was defined as that of the density matrix describing the statistics of measurements of the geometry enclosed by such a boundary geometry. This definition of the entropy can be extended to larger local regions on $k$-CH graphs.

We analyzed a concrete example involving the superposition of automorphism-invariant states on two distinct $2$-CH graphs. We considered the superposition of Bell-network states with fixed spins on graphs $\Gamma_5$ and $\Gamma_6$ with $5$ and $6$ four-valent nodes, respectively. We set all spins $j_\ell=1/2$, for simplicity, and computed the entropy in terms of the probability $p$ associated with the graph $\Gamma_5$, both for a single node and for two adjacent nodes. The entropy of the superposition can exceed that of the individual states, as could be expected, opening the possibility of fixing the amplitudes of the  states in the superposition by imposing a maximization of the entropy. It would be interesting to study more general superpositions involving larger spins and more complex graphs in order to determine the consequences of a condition of maximal entropy for superpositions of graphs.

The entropy of a local region was defined in this work for a cosmological state, but it should be possible to extend the strategy adopted for that purpose to more general circumstances, presenting another natural direction for the extension of our results. We expect that the entropy of a local region specified by a boundary geometry can be likewise defined for any state of the geometry, and more general cases than those explored here can be studied elsewhere.


\begin{acknowledgments}
N.Y. acknowledges financial support from the Conselho Nacional de Desenvolvimento Cient\'ifico e Tecnol\'ogico (CNPq) under Grant No. 306744/2018-0. B.B. acknowledges support from the National Science Foundation of China (NSFC) by Grants Nos. 11875006 and 11961131013.
\end{acknowledgments}

\appendix

\section{A list of useful graphs, including nomenclature, definitions and basic features}
\label{sec:graph-list}

General references for the definitions below are~\cite{bondy08,harary72, coxeter, coxeter2, coxeter22}.

\paragraph{Cycle graph $C_{N}$:} A graph with $N$ nodes comprising a single cycle through all nodes. The number of nodes in $C_{N}$ equals the number of links, and every node has degree two. 

\paragraph{Complete graph $K_{N}$:} A simple graph with the property that each pair of distinct nodes is connected by a link (see Fig.~\ref{fig:pentagram}). There is only one complete graph with $N$ nodes up to a graph isomorphism. The size of a complete graph is given by $L = N (N+1)/2$. 

\paragraph{Complete multipartite graph $K(r,s)$ of $r$ parts of size $s$:} The node set is the union of $r$ sets $X_1,\dots,X_r$ of order $|X_i|=s$. Two nodes are adjacent if and only if they belong to distinct sets $X_i$ and $X_j$, $i \neq j$. 

\paragraph{Complete bipartite graph $K_{N,M}$} A graph whose node set can be partitioned into two subsets $M$ and $N$ such that no link has both endpoints in the same subset, and any point in $M$ is adjacent to any point in $N$.

\paragraph{Line graph $\mathrm{L}(\Gamma)$ of a graph $\Gamma$:} The line graph $\mathrm{L}(\Gamma)$ of a graph $\Gamma$ has as node set the link set of $\Gamma$. Two nodes in the line graph are joined if and only if they correspond to links sharing a common node in the original graph.

\paragraph{Cayley graph:} A (uncolored and undirected) Cayley graph $\Gamma(G,S)$ is labelled by a group $G$ and a generating set $S$ of $G$. The node set of the Cayley graph is the group $G$. For any $g \in G$ and $s \in S$, the nodes $g$ and $g s$ are joined by a link, i.e., the link set is formed by pairs $\{g, g s\}$. 

\paragraph{Tree graph $T_r$:} A tree is a connected graph without cycles. A tree is regular if all vertices have the same degree. We denote by $T_r$ the regular tree of valency $r$.

\paragraph{Schl\"{a}fli symbol:} Schl\"{a}fli symbols are finite lists $\{p,q,r,...\}$ of natural numbers used to represent regular polytopes and tessellations of spheres, hyperbolic and Euclidean spaces. A Schl\"{a}ffi symbol with a single index $\{p\}$ represents a $p$-sided regular polygon. A Schl\"{a}fli symbol $\{p,q\}$ with two indices represents the object formed by $p$-sided regular polygons glued so that $q$ polygons meet at each node. Put $\mu=(p-2)(q-2)-4$, with $p,q>2$. The symbols $\{p,q\}$ describe tessellations of the Euclidean plane ($\mu=0$), sphere ($\mu < 0$) and the hyperbolic plane ($\mu > 0$) \cite{coxeter,coxeter-whitrow}. There is a finite number of solutions for the spherical case (the Platonic solids) and for the plane case (triangular lattice: $\{3,6\}$, square lattice: $\{4,4\}$ and hexagonal honeycomb $\{6,3\}$), but an infinite number of solutions for the hyperbolic case. Similarly, a Schl\"{a}fli symbol $\{p,q,r\}$ with three indices describes the object formed by gluing regular polytopes $\{p,q\}$ so that $r$ such polytopes meet at each link. As in the two-dimensional case, the type of curvature of the embedding space is determined by $\mu = \sin(\pi/p) \sin(\pi/r) - \cos(\pi/q)$, which gives Euclidean ($\mu =0$), spherical ($\mu > 0$) and hyperbolic ($\mu < 0$) 3d spaces, respectively \cite{coxeter,coxeter-whitrow}. For instance, $\{4,3,4\}$ defines cubes as building blocks, with four of them around each link, describing a cubic lattice.

\paragraph{Honeycomb:} A honeycomb is a tessellation of an $n$-dimensional Euclidean space with polyhedral or higher-dimensional cells. A hyperbolic honeycomb is a tessellation of an $n$-dimensional hyperbolic space with hyperbolic polyhedra or higher-dimensional cells. A honeycomb is called regular if the group of isometries preserving the tessellation acts transitively on all the elements of the honeycomb (vertices, edges, faces and cells). The cubic lattice is the only regular honeycomb in 3d Euclidean space. The dual graph of an infinite regular honeycomb with Schl\"{a}fli symbol $\{p,q,r\}$ is denoted by $\mathrm{Hc}(\{r,q,p\})$.

\paragraph{Lattice $\mathcal{L}_{n}^{(\mathrm{V})}$:} An $n$-dimensional lattice $\mathcal{L}_{n}^{(\mathrm{V})}$ in $\mathds{R}^n$ is a graph with valency $V$ whose node set is a subgroup of the additive group $\mathds{R}^n$ that is isomorphic to the additive group $\mathds{Z}^n$. For example, for any basis of $\mathds{R}^n$, the subgroup of all linear combinations of the basis vectors with integer coefficients forms a lattice $\mathcal{L}_{n}^{(\mathrm{2n})}$. A triangular grid, a tessellation of the plane formed by identical equilateral triangles, is a lattice $\mathcal{L}_{2}^{(\mathrm{6})}$.

\paragraph{Circulant graph $\mathrm{Ci}_{N}$:} A circulant graph is a graph with $N$ nodes $n_i, \, i=0,\dots,N-1$, such that if $n_j$ and $n_{(j+d) \, \mathrm{mod} \, N}$ are adjacent, then the nodes $n_{j'}$ and $n_{(j'+d) \, \mathrm{mod} \, N}$ are also adjacent, for all $j'$. The automorphism group of the graph includes all cyclic permutations of its nodes.

\paragraph{Hypercube graph $Q_n$:} A hypercube graph with valency $n$ is the graph formed by the nodes and links of an $n$-dimensional hypercube. It has $2^n$ nodes and $2^{n-1} \, n$ links.

\paragraph{Perfect matching:} A matching in a graph is a set of pairwise non-intersecting links. A perfect matching is a matching for which every node of the graph is in some link of the matching. A perfect matching has $N/2$ links, implying that perfect matchings only exist for graphs with an even number of nodes.

\paragraph{Rado graph:} The nodes of the Rado graph are the natural numbers $\mathbb{N}=\{0,1,2,\dots\}$. For $i < j$, the nodes $i$ and $j$ are joined if and only if the $i$-th digit of $j$ (in base 2) is 1. The Rado graph is a self-complementary graph, i.e., it is isomorphic to its complement. The valency of the Rado graph is infinite.

\paragraph{Henson graphs:} The Henson graph $H_N$ is the unique countable infinite homogeneous graph that does not contain a complete graph $K_{N}$ but contains all $K_{N}$-free finite graphs as induced subgraphs. For instance, $H_3$ is a triangle-free graph (no three vertices form a triangle of links) that contains all finite triangle-free graphs. The valency of the Henson graph is infinite.

\paragraph{Petersen graph $O_3$:} The Petersen graph is the complement of the line graph of $K_5$. It has $10$ nodes with valency $V=3$ and $15$ links. See Fig.~\ref{fig:CHgraphs}.

\paragraph{Clebsch graph $\Box_5$:} The Clebsch graph has 16 nodes with valency 5 and 40 links. It can be constructed from a 5-dimensional hypercube graph $Q_5$ by identifying every pair of opposite nodes. See Fig.~\ref{fig:CHgraphs}.


\end{document}